\documentclass{article}

\usepackage{epsfig}
\def\ligne#1{\hbox to \hsize{#1}}
\def\PlacerEn#1 #2 #3 {\rlap{\kern#1\raise#2\hbox{#3}}}
\def\cqfd{\hbox{\kern 2pt\vrule height 6pt depth 2pt width 8pt\kern 1pt}}
\def\leurre{\noindent\leftskip0pt\small\baselineskip 10pt}

\newtheorem{thm}{Theorem}

\newtheorem{fig}{Figure}
\newtheorem{tab}{Table}

\font\ttix=cmtt9

\def\grostrait{\ligne{\vrule height 1pt depth 1pt width \hsize}}
\def\demitrait{\ligne{\vrule height 0.5pt depth 0.5pt width \hsize}}

\begin{document}
\begin{center}
{\bf \Large A new weakly universal cellular automaton in the $3D$~hyperbolic space
with two states}
\vskip 5pt
Maurice {\sc Margenstern}
\vskip 2pt
Universit\'e Paul Verlaine $-$ Metz, IUT de Metz,\\
LITA EA 3097, UFR MIM,\\
Campus du Saulcy,\\
57045 METZ C\'edex 1, FRANCE\\
{\it e-mail}: {\tt margens@univ-metz.fr}\\
{\it Web page}: {\tt http://www.lita.sciences.univ-metz.fr/\~{}margens}
\end{center}
\vskip 5pt
{\parindent 0pt\leftskip 20pt\rightskip 20pt
{\bf Abstract} $-$ In this paper, we show a construction of a weakly universal cellular
automaton in the~$3D$ hyperbolic space with two states. Moreover, based on a new 
implementation of a railway circuit in the dodecagrid,the construction is a truely $3D$-one.
\par}

\section{Introduction}
\label{intro}
   In this paper, we construct a weakly universal cellular automaton in 
the~$3D$ hyperbolic space with two states. Moreover, based on a new
implementation of a railway circuit in the dodecagrid,the construction is a truely $3D$-one.

   The dodecagrid is the tiling~$\{5,3,4\}$ of the $3D$~hyperbolic space, and we
refer the reader to~\cite{mmgsFI,mmbook1} for an algorithmic approach to  this
tiling. We also refer the reader to \cite{mmJCA3D,mmbook2} for an implementation
of a railway circuit in the dodecagrid which yields a weakly universal cellular
automaton with 5~states. The circuit is the one used in other papers by the author,
alone or with co-authors, inspired by the circuit devised by Ian Stewart,
see~\cite{stewart}. The notion of weak universality is discussed in previous papers,
see for instance~\cite{woods,mmTCSfbdu,mm2tur,mmJCA3D} and comes from the
fact that the initial configuration is infinite. Howerver, it is not an arbitrary
configuration: it has to be regular at large according to what was done previously,
see~\cite{mmJCA3D,mmsyPPL,mmsyENTCS,mmTCS4}.

   In~\cite{mmarXiv1CAtoHCA}, I found a way to implement $1D$~cellular automata in
two grids of the hyperbolic plane and one grid of the $3D$~hyperbolic space: the pentagrid,
the heptagrid and the dodecagrid respectively. In this paper, I proved that such an 
implementation is possible without increasing the number of states of the automaton which
is implemented when this is performed in the heptagrid or in the dodecagrid. I proved
that such an implementation is still possible in the pentagrid when the cellular automaton
satisfies an additional condition. It turns out that rule~110 of elementary cellular 
automata, see for instance~\cite{cook,wolfram}, satisfies this condition. This proves that
there are weakly universal cellular automata with two states, in the pentagrid, in the
heptagrid and in the dodecagrid.

   However, such a construction is not fully satisfying for the reason that it does not
make use of the geometrical properties of the plane or the space. Also, the result
is obtained by application of a very strong result on $1D$~cellular automata, the weak
universality of rule~110, which requires very huge configurations and enormous computations.
In this paper, we provide a construction which is far more simpler and requires very
few resources for verification.

   As we definitely changed the implementation of the railway circuit, we pay a new visit
to the description of the circuit in Section~\ref{railway} as well as to the
previous implementations in Section~\ref{previously}. In Section~\ref{new}, we
define the elements of the new implementation and, in Section~\ref{scenario} we describe 
the scenario of the simulation. In Section~\ref{the_rules}, we give the rules of the
automaton and we give a brief account of the computer program which checked the rules and
the simulation, allowing us to prove:

\begin{thm}\label{weakuniv2}
There is a weakly universal cellular automaton in the dodecagrid which is weakly 
universal and which has two states exactly, one state being the quiescent state. 
Moreover, the cellular automaton is rotation invariant and the set of its cells
changing their state is a truely $3D$-structure. 
\end{thm}

In Section~\ref{conclusion} we look at the remaining tasks.

\section{The railway circuit}
\label{railway}
As initially devised in~\cite{stewart} and then mentioned
in~\cite{mmCSJMtrain,fhmmTCS,mmsyPPL,mmsyENTCS,mmbook2},
the circuit uses tracks represented by lines and quarters of circles and switches.
There are three kinds of switches: the {\bf fixed}, the {\bf memory} and the
{\bf flip-flop} switches. They are represented by the schemes given in
Fig.~\ref{aiguillages}.

\vskip 20pt
\vtop{
\vspace{-10pt}
\setbox110=\hbox{\epsfig{file=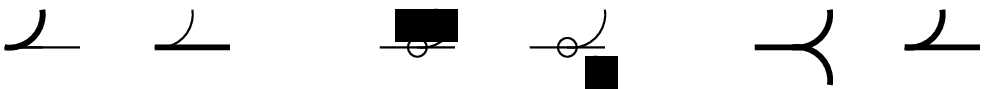,width=280pt}}
\ligne{\hfill
\PlacerEn {-305pt} {0pt} \box110
}
\vspace{-5pt}
\begin{fig}
\label{aiguillages}
\leurre
The three kinds of switches. From left to right: fixed, flip-flop and memory switches.
\end{fig}
}
\vskip 10pt
   Note that a switch is an oriented structure: on one side, it has a single
track~$u$ and, on the the other side, it has two tracks~$a$ and~$b$. This
defines two ways of crossing a switch. Call the way from~$u$ to~$a$ or~$b$
{\bf active}. Call the other way, from~$a$ or~$b$ to~$u$ {\bf passive}. The
names comes from the fact that in a passive way, the switch plays no role on
the trajectory of the locomotive. On the contrary, in an active
crossing, the switch indicates which track, either~$a$ or~$b$ will be followed by
the locomotive after running on~$u$: the new track is called the {\bf selected}
track.

   As indicated by its name, the {\bf fixed switch} is left unchanged by the
passage of the locomotive. It always remains in the same position: when
actively crossed by the locomotive, the switch always sends it onto the same
track. The flip-flop switch is assumed to be crossed actively only. Now,
after each crossing by the locomotive, it changes the selected track.
The memory switch can be crossed by the locomotive actively and passively.
In an active passage, the locomotive is sent onto the selected track. Now, the
selected track is defined by the track of the last passive crossing by the
locomotive. Of course, at initial time, the selected track is fixed.


\vskip 10pt
\vtop{
\vspace{-10pt}
\setbox110=\hbox{\epsfig{file=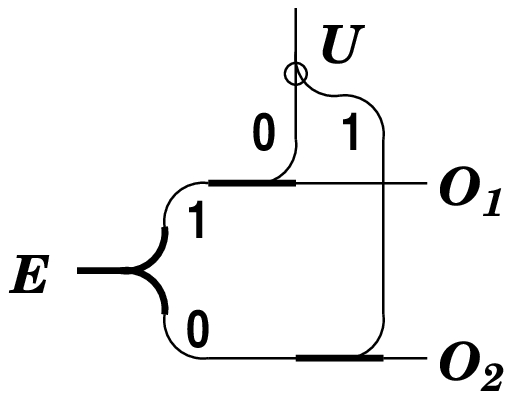,width=200pt}}
\ligne{\hfill
\PlacerEn {-285pt} {0pt} \box110
}
\vspace{-15pt}
\begin{fig}
\label{element}
\leurre
The elementary circuit.
\end{fig}
}
\vskip 10pt

   With the help of these three kinds of switches, we define an
{\bf elementary circuit} as in~\cite{stewart}, which exactly contains one bit of
information. The circuit is illustrated by Fig.~\ref{element}, below and it is
implemented in the Euclidean plane. It can be remarked that the working of the 
circuit strongly depends on how the locomotive enters it. If the locomotive enters 
the circuit through~$E$,
it leaves the circuit through~$O_1$ or~$O_2$, depending on the selected track
of the memory switch which stands near~$E$. If the locomotive enters through~$U$,
the application of the given definitions shows that the selected track at the
switches near~$E$ and~$U$ are both changed: the switch at~$U$ is a flip-flop
which is changed by the actual active passage of the locomotive and the switch
at~$E$ is a memory one which is changed because it is passively crossed by the
locomotive and through the non-selected track. The actions of the locomotive
just described correspond to a {\bf read} and a {\bf write} operation on the
bit contained by the circuit which consists of the configurations of the
switches at~$E$ and at~$U$. It is assumed that the write operation is triggered
when we know that we have to change the bit which we wish to rewrite.

   From this element, it is easy to devise circuits which represent different
parts of a register machine. As an example, Fig.~\ref{unit} illustrates
an implementation of a unit of a register.

Other parts of the needed circuitry are described
in~\cite{mmCSJMtrain,fhmmTCS}. The main idea in these different parts is
to organize the circuit in possibly visiting several elementary circuits
which represent the bits of a configuration which allow the whole system
to remember the last visit  of the locomotive. The use of this technique is
needed for the following two operations.

When the locomotive arrives to a register~$R$, it arrives either to
increment~$R$ or to decrement it. As can be seen on Fig.~\ref{unit}, when the
instruction is performed, 
\ligne{\hfill}
\vskip 10pt
\vtop{
\vspace{-10pt}
\setbox110=\hbox{\epsfig{file=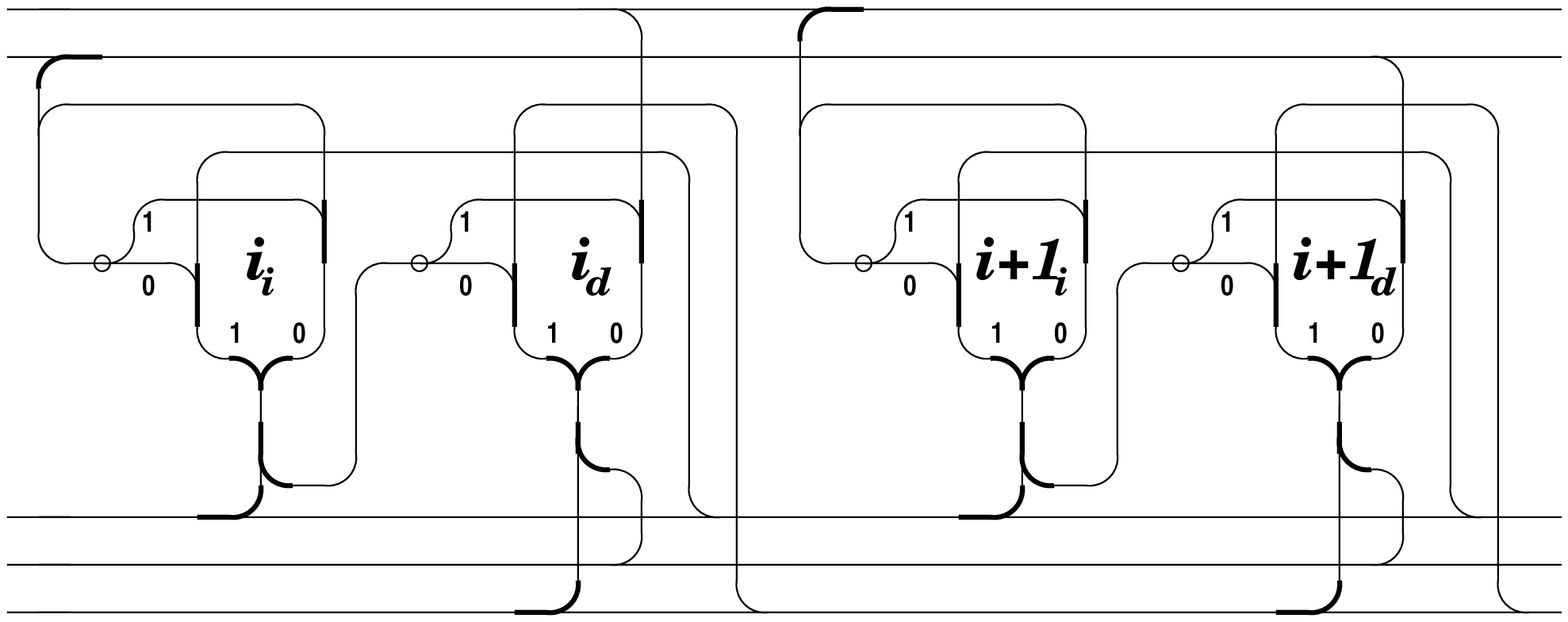,width=320pt}
\PlacerEn {-315pt} {24pt} {\small $i$}
\PlacerEn {-315pt} {14.5pt} {\small $d$}
\PlacerEn {-315pt} {5pt} {\small $r$}
\PlacerEn {-315pt} {127pt} {\small $j_1$}
\PlacerEn {-315pt} {117pt} {\small $j_2$}
}
\ligne{\hfill\hskip 20pt
\PlacerEn {-325pt} {0pt} \box110
\hskip 10pt}
\begin{fig}
\label{unit}
\leurre
Here, we have two consecutive units of a register. A register contains
infinitely many copies of units. Note the tracks $i$, $d$, $r$, $j_1$ and~$j_2$.
For incrementing, the locomotive arrives at a unit through~$i$ and it leaves the
unit through~$r$. For decrementing, it arrives though~$d$ and it leaves
also through~$r$ if decrementing the register was possible, otherwise, it leaves
through~$j_1$ or~$j_2$.
\end{fig}
}

\vskip 10pt
\vtop{
\vspace{-10pt}
\setbox110=\hbox{\epsfig{file=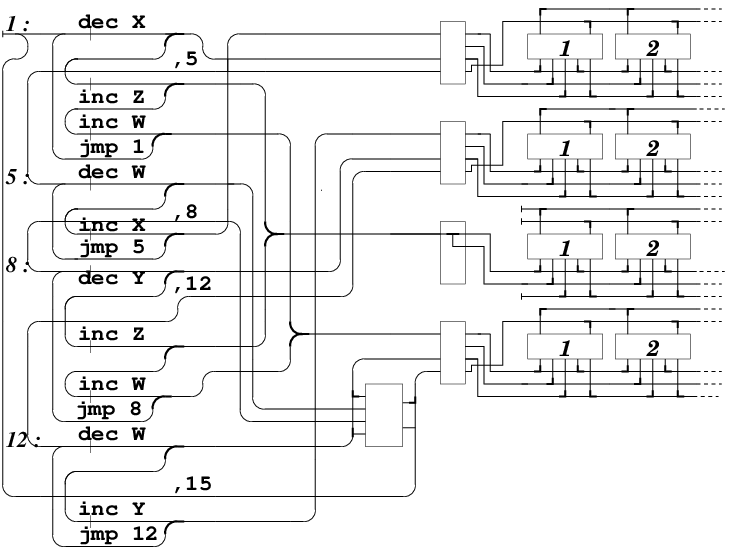,width=280pt}}
\ligne{\hfill
\PlacerEn {-325pt} {0pt} \box110
}
\vskip-15pt
\begin{fig}
\label{example}
\leurre
An example of the implementation of a small program of a register machine.
On the left-hand side of the figure, the part of the sequencer. It can be noticed
how the tracks are attached to each instruction of the program. Note that there
are four decrementing instructions for~$W$: this is why a selector gathers
the arriving tracks before sending the locomotive to the
control of the register. On the way back, the locomotive is sent on the right
track.
\end{fig}
}
\vskip 10pt
\noindent
the locomotive goes back from the register by the
same track. Accordingly, we 
need somewhere to keep track of the fact whether
the locomotive incremented~$R$ or it decremented~$R$. This is one type of control.
The other type comes from the fact that several instructions usually apply
to the same register. Again, when the locomotive goes back from~$R$,
in general it goes back to perform a new instruction which depends on the one
it has just performed on~$R$. Again this can be controlled by what we called
the {\bf selector} in~\cite{mmCSJMtrain,fhmmTCS}.

At last, the dispatching of the locomotive on the right track for the next
instruction is performed by the {\bf sequencer}, a circuit whose main structure
looks like its implementation in the classical models of cellular automata such
as the game of life or the billiard ball model. The reader is referred to the
already quoted papers for full details on the circuit. Remember that this
implementation is performed in the Euclidean plane, as clear from
Fig.~\ref{example} which illustrates the case of a few lines of a program of
a register machine.

   Now, we turn to the implementation in the hyperbolic plane. We refer the reader
to~\cite{mmbook1,mmarXiv1CAtoHCA} for the few required features of hyperbolic geometry.
As announced in the introduction, first we look at the previous implementations.

\section{The previous hyperbolic implementations}
\label{previously}

   The first implementations were based on the following idea. The tracks are
realized by a $1D$-structure and the switches are realized by the meeting of
the tracks at a cell. There were no other ingredients and this explains why
the first obtained automaton had 22~states, see~\cite{fhmmTCS}. In this 
paper, we indicated the implementation of the tracks and of the circuit
with many details. These details are reproduced and completed in~\cite{mmbook2}
from which Figures~\ref{hca_cross_paths}, \ref{unit_in_Hii}, \ref{h_parallel},
\ref{unit_block} and~\ref{hca_global} are taken.

\setbox115=\hbox{\epsfig{file=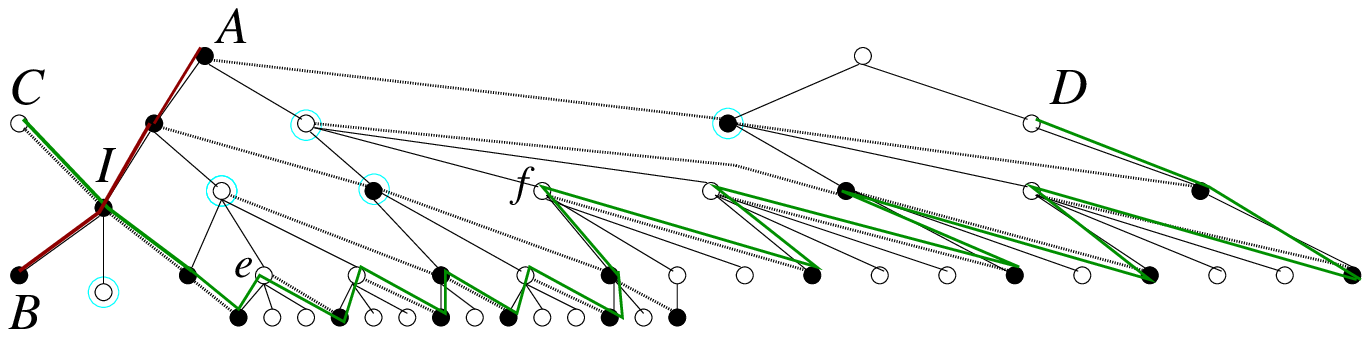,width=330pt}}
\vskip 14pt
\vtop{
\ligne{\hfill
\PlacerEn {-340pt} {0pt} {\box115}
}
\vskip -5pt
\begin{fig}\label{hca_cross_paths}
\leurre
Crossing of paths: the basic pattern.
\vskip2pt
Note the nodes which are circled. Both paths must avoid these nodes in order
to observe the following rule: a node coloured with~$B$ has at most two neighbours,
also coloured with~$B$. This condition entails the shape of the basic
pattern~$ef$.
\end{fig}
}

  Thanks to the $3D$~representation, as this was already noticed in~\cite{mmJCA3D},
we can avoid crossings and so, we have no more to bother with the complications
which arise from this situation as witnessed by Figure~\ref{hca_cross_paths}. However,
the other figures remain valid and, as witnessed by Figure~\ref{h_parallel}, we can
see that complicated patterns cannot be avoided.

   Now, as Figures~\ref{unit_block} and~\ref{hca_global} suggest, we can define
an analogue of the segments of lines and the quarters of circles used in the
Euclidean implementation. Here, we replace the segments of lines by what we called 
{\bf vertical} in~\cite{mmsyENTCS,mmarXiv3} in different tilings, and we replace
the quarter of circles by what we called {\bf horizontal} in the same papers.

\setbox115=\hbox{\epsfig{file=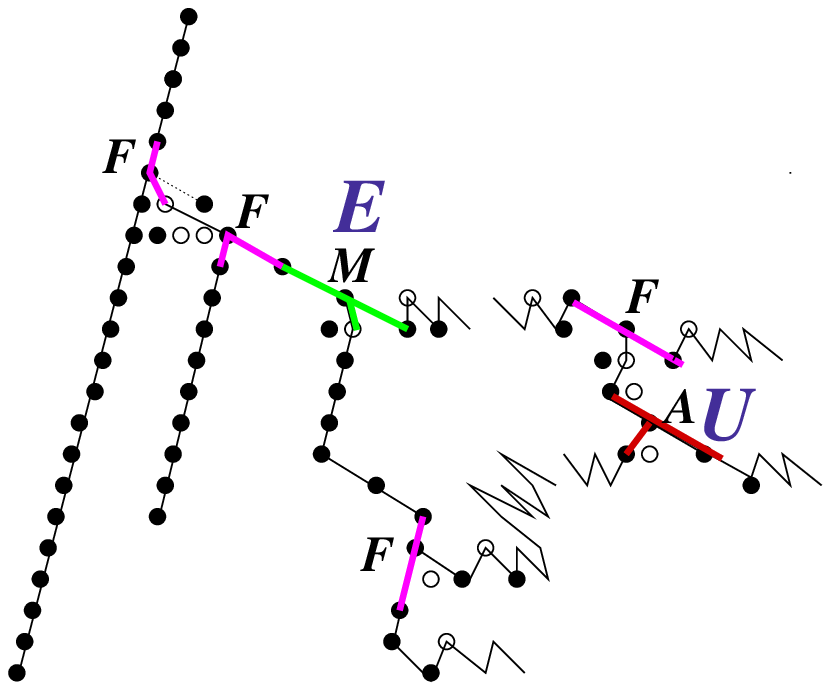,width=240pt}}
\vtop{
\ligne{\hfill
\PlacerEn {-310pt} {-215pt} {\box115}
}
\begin{fig}\label{unit_in_Hii}
\leurre
Implementation of the elementary unit in the Fibonacci tree.
\vskip 2pt
Note the implementation of the three kinds of switches. Also note the
implementation of the various switches, depending on the desired direction of the
path: the additional nodes represent the exact neighbours of the intersection
node which the paths exactly go through.
\end{fig}
}

\setbox110=\hbox{\epsfig{file=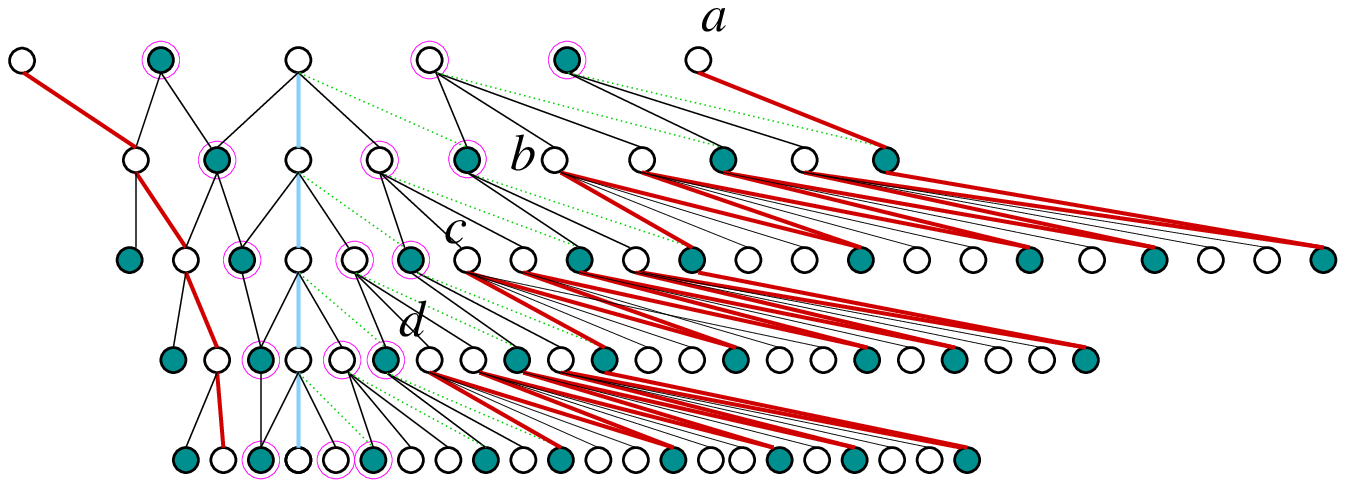,width=300pt}}
\vskip 30pt
\vtop{
\ligne{\hfill
\PlacerEn {-320pt} {0pt} {\box110}
}
\begin{fig}\label{h_parallel}
\leurre
Implementation of parallel paths in the Fibonacci tree.
\vskip 2pt
In the figure, forbidden nodes are circled. Note the reproduction of
the shape of the broken line~$ab$ on the broken lines~$bc$ and~$bd$. The
reproduction is the result of a simple shift.
\end{fig}
}

   In fact, the implementation of~\cite{mmarXiv3} has taken benefit of a new implementation
of horizontals with respect to~\cite{mmsyENTCS}. 

   Already in~\cite{mmJCA3D}, the idea of marking the intersection point of three tracks
meeting at a switch by a colour corresponding to the type of switch was replaced by
signals placed around the considered cell. This was easy in the $3D$~space, which explains 
that there we first obtained a weakly universal cellular automaton with 5~states. Moreover,
the rules of this automaton satisfy a very strong property: they are not only rotation
invariant, they are also different of each other by the Parikh vector associated to
the rules. This vector is obtained by taking each state considered as a letter, and to
write the rule as a word consisting of blocks of the same letter, each letter being repeated
as many times as it appears among the neighbours of the cell. We said that the rules
were pairwise {\bf lexicographically} different.

\setbox110=\hbox{\epsfig{file=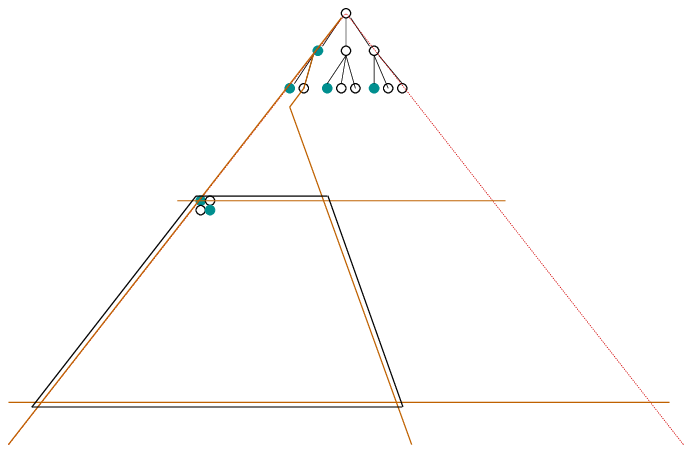,width=240pt}}
\vtop{
\ligne{\hfill
\PlacerEn {-320pt} {0pt} {\box110}
}
\begin{fig}\label{unit_block}
\leurre
Delimitation of a block in a Fibonacci tree~$F$.
\vskip 2pt
In the figure, we represent the levels between which the block is delimited.
The branch which delimits the block on its left-hand side is supported by
the leftmost branch of~$F$. The branch which delimits the right-hand side
of the block is supported by the rightmost branch of a sub-tree of~$F$.
\end{fig}
}

In \cite{mmTCS4}, a new idea is introduced in the context of the heptagrid. The
idea consists in replacing the linear structure of the tracks by a new one:
the orbit of the locomotive remains a $1D$-structure, but it is not materialized by
cells which are different from the quiescent state. The track are materialized
by cells which are immediate neighbours of this orbit. This allowed to obtain
a weakly cellular automaton in the heptagrid with 4~states only, the rules being also
rotation invariant. Moreover, the implementation is a true planar one: there are a
lot of cycles in the orbit of the locomotive. These immediate neighbours of the orbit 
were called {\bf milestones} in~\cite{mmTCS4}. The milestone technique was used 
again in~\cite{mmarXiv3}, which allowed to obtain a weakly universal cellular with
three states only, the rules being again rotation invariant. However, the pairwise
lexicographic difference could not be kept.

   Now, \cite{mmarXiv3} introduces a new representation with respect 
to~\cite{mmJCA3D}. Indeed, remark that the trace of the dodecagrid on a plane~$\Pi_0$
which is supported by a face of one of the dodecahedra of the tiling is a copy
of the pentagrid. Then, by using the Schlegel diagrams used in~\cite{mmJCA3D},
we can project each dodecahedron which is in contact with~$\Pi_0$ onto~$\Pi_0$ itself.
This allows us to obtain a representation which is illustrated by the figures of the 
next section.

\setbox110=\hbox{\epsfig{file=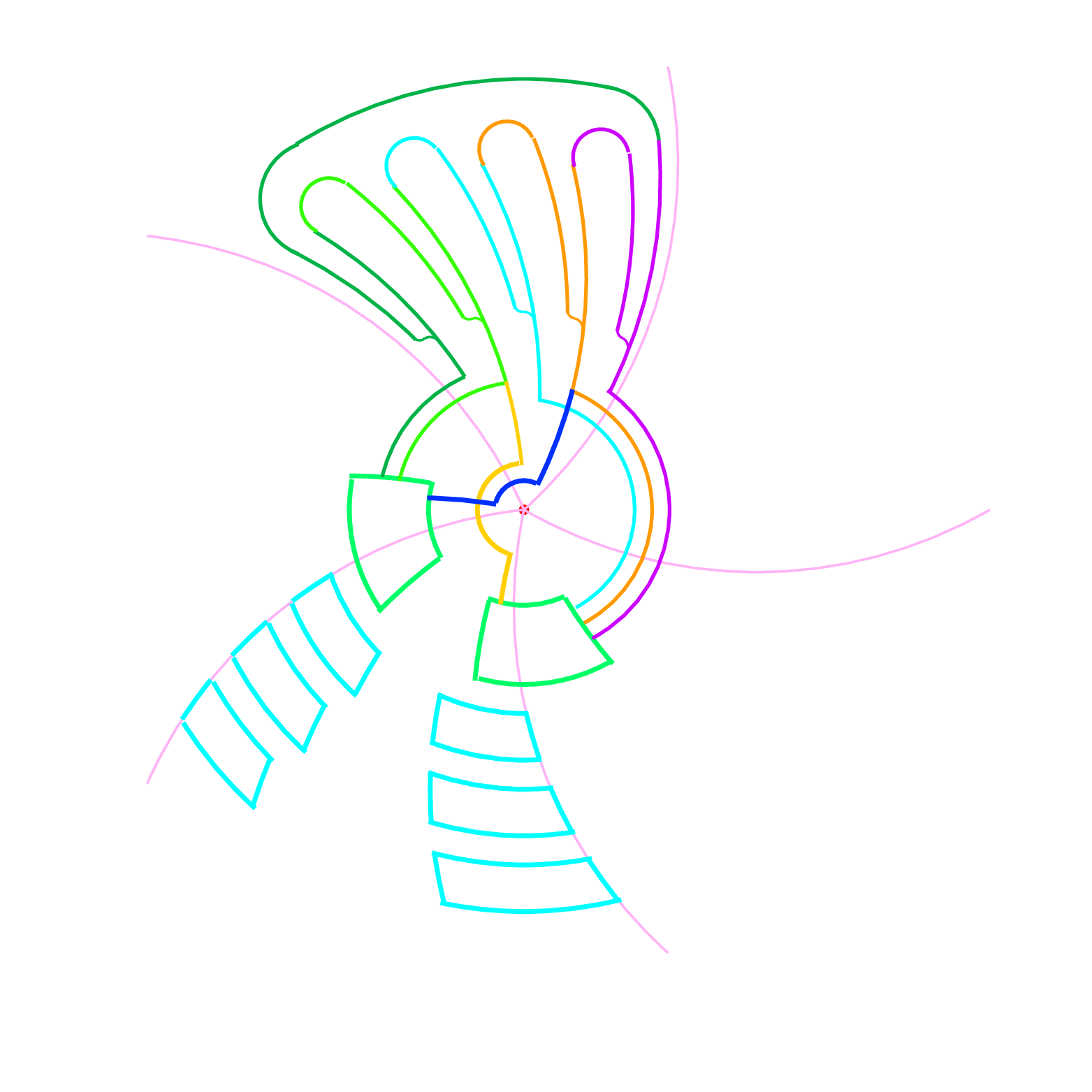,width=280pt}}
\vtop{
\ligne{\hfill
\PlacerEn {-320pt} {0pt} {\box110}
}
\vspace{-40pt}
\begin{fig}\label{hca_global}
\leurre
Schematic representation of the whole circuit which corresponds to 
Figure~{\rm\ref{example}}.
\vskip 2pt
In the figure, register~$1$ is on the leftmost branch while register~$2$ is
on the rightmost one. The instructions which increment register~$2$
are in orange, while the instruction which decrements it is in purple. Similarly,
for register~$1$: the incrementing instructions is in dark green and the
decrementing one is in light green. The return track which
corresponds to jumps of decrementing instructions are in yellow for register~$2$,
in dark blue for register~$1$.
\end{fig}
}

   It is probably the place to introduce here the convention which is used with this
representation: if not otherwise mentioned, in this projection, a face~$F$ of a 
dodecahedron~$D$ takes the colour of the state taken by the neighbour of~$D$ which shares
with it the face~$F$. 
 
\section{A new implementation in the dodecagrid}
\label{new}

   Now, we start the new implementation by a new analysis of the railway circuit.

This new implementation was motivated by the remark that the motion rules of the locomotive
work also if the locomotive is reduced to one cell which is of the same colour of the
milestones: the black colour as now we have only two colours, white and black. But they
work in one direction only. The same pattern cannot be used for moving the locomotive,
which we call from now on {\bf particle}, in both directions. One direction must be fixed
for this pattern, the other direction being forbidden. 

The main idea is to introduce tracks by pairs, as this is mostly the case in real life. 
There is a track for one direction and another one for the opposite direction. Now, we can
see in Figure~\ref{h_parallel} how complex is the construction of a 'parallel` line in 
the hyperbolic plane. We shall see later that thanks to the third dimension, we can
find a more efficient solution.

   Now, before turning to this solution, we have to look at how the switches work
in this new setting. 

\subsection{The new switches}
\label{new_switches}

   For this analysis, we shall again use the tracks~$u$, $a$ and~$b$ defined in
Sectin~\ref{railway}. We remind that the active passage goes from $u$~to either~$a$
or~$b$ and that the passive crossing goes from either~$a$ or~$b$ to~$u$. As we split
the ways into two tracks, we shall denote them by $u_d$, $u_r$, $a_d$, $a_r$, $b_d$ and
$b_r$ respectively, where the subscript $d$~indicates the active direction and $r$~indicates
the return one. {\it A priori}, this defines two switches: the first one from $u_d$ 
to~$a_d$ or~$b_d$ and the second one from $a_r$ or~$b_r$ to$u_r$. We shall call the 
first one the {\bf active switch} and the second one the {\bf passive switch}.
Note that each of these new switches deals with one-way tracks only. This can be
illustrated by the right-hand side picture of Figure~\ref{switches}.

   As the flip-flop switch is used in an active passage only,
it makes use of single tracks only, in the direction to the switch for the way~$u$
defined in Section~\ref{railway}, in the direction coming from the switch on the two 
ways~$a$ and~$b$ defined in Section~\ref{railway} too. Accordingly, the flip-flop
switch is only an active switch. There is no passive switch in this case.

\setbox110=\hbox{\epsfig{file=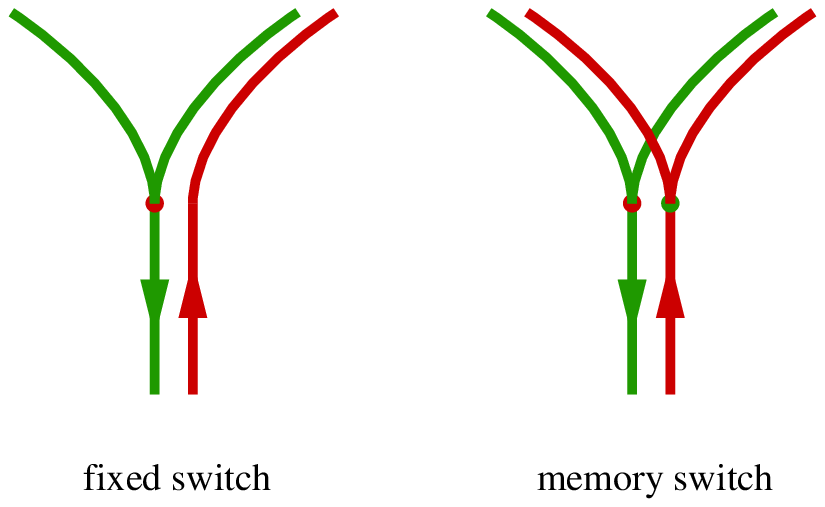,width=280pt}}
\vtop{
\ligne{\hfill
\PlacerEn {-290pt} {0pt} {\box110}
}
\vspace{-40pt}
\begin{fig}\label{switches}
\leurre
The new switches. On the left-hand side, the fixed switch;
on the right-hand side: the memory switch.
\end{fig}
}

   And so, we remain with the fixed and the memory switches. 

   First, let us look at the fixed switch. As the switch is fixed, we may assume
that $u$ always~goes to~$a$. This means that the track $b_d$ is useless. Now,
it is plain that $u_d$ and~$a_d$ sonstitute the two rays of a line issued from a
point of the line. Consequently, there is no active fixed switch. The switch is 
concerned by the return tracks only: $u_r$, $a_r$ and~$b_r$, it is a passive switch. 
It is easy to see that it works as a {\bf collector}: it collects what comes from 
both $a_r$ and~$b_r$ and send this onto~$u_r$. Now, as at any time there is at most 
one particle arriving at the switch from the union of~$a_r$ and~$b_r$, the collector 
receives the particle from~$a_r$ or~$b_r$ alternately, never at
the same time, see the left-hand side picture of Figure~\ref{switches}. 

   Now, let us look at the memory switch. It is clear that, in this case,
we have both an active and a passive switch. However, a closer look at the situation
shows that the passive memory switch has a tight connection with the fixed switch
and that the active memory switch has a tight connection with the flip-flop switch.
We shall return to this point in Section~\ref{scenario}.
 
   For the moment, we go back to the implmentation of the tracks in the dodecagrid.

\subsection{Implementing two tracks}
\label{implement_tracks}

   As already announced, we can exploit the third dimension in order to get a more 
efficient way. We first take the same general principle as in~\cite{mmarXiv3}. This means
that a vertical segment is defined by a line~$\ell$ of the pentagrid which is a line which 
supports a side of a pentagon of the tiling. We say that $\ell$~is the {\bf guideline}
of the vertical segment and it consists of a sequence of sides. Now, in~\cite{mmarXiv3}, 
we fixed a face of a dodecahedron which fixes a plane~$\Pi_0$. As already mentioned, 
$\Pi_0$ consists of pentagons which are faces of dodecahedra of the tiling and these 
faces realize a copy of the pentagrid on~$\Pi_0$. In~\cite{mmarXiv3} a vertical segments
is a sequence of dodecahedra which follows the guideline and which have a face on~$\Pi_0$
in a such way that all the dodecahedra of the vertical segment lie in the same half-space 
defined by~$\Pi_0$. We say that the vertical segment is {\bf above}~$\Pi_0$. We can
see such a vertical line in Figure~\ref{dodec_vert}, realized by the yellow tiles which
can be seen there. We can see in the figure that each dodecahedron which is above~$\Pi_0$
is projected as a Schlegel diagram on the face which lies on~$\Pi_0$.

   Our implementation for vertical lines is based on the one we did in~\cite{mmarXiv3}.
Remark that, in this implementation, the dodecahedra of the line have a milestone
which is below~$\Pi_0$. We can see three milestones on the projection of each
dodecahedron, but as a dodecahedron of a vertical line has four milestones around itself,
the fourth one is the neighbour which can be seen from the face which lies on~$\Pi_0$.
Fix $\Delta$ a dodecahedron of the vertical segment~$V$ and let $F$~be its face on~$\Pi_0$.
Denote by~$[F]$ the reflection of~$\Delta$ in~$F$. We know that $[F]$~is a milestone
of~$\Delta$. As $\Delta$ is above~$\Pi_0$, $[F]$~is below~$\Pi_0$. Now, consider
the face~$G$ of~$\Delta$ which has a side on the guideline and which is not~$F$. We
can see that $[G]$~is also a milestone which has a face~$H$ on~$\Pi_0$, so that $G$ is
also above~$Pi_0$. As there are four dodecahedra around an edge in the dodecagrid,
the four dodechedra which are around the side~$s$ of the guideline shared by~$F$ and~$G$
are $\Delta$, $[F]$, $[G]$ and~$[H]$. Now, we can see that $\Delta$ and~$[H]$ are the
reflection of each other under the reflection in~$s$. This reflection also shows that
the situation of~$[F]$ and~$[G]$ with respect to~$\Delta$ is the same as their situation
with respect to~$[H]$. Let us call~$[F]$ the {\bf ballast} of~$\Delta$. Then, we can
see that $[G]$ can be considered as the ballast of~$[H]$. The reflection in~$s$ transforms
$\Delta$ into~$[H]$ and the milestones of~$\Delta$ into milestones of~$[H]$. This can be seen
in the below picture of the right-hand side of Figure~\ref{dodec_vert}. The
reflection of~$V$ in~$s$ defines a vertical line~$W$ which is this time 
{\bf below}~$\Pi_0$. But the directions of~$V$ and~$W$ are the same. So that
keeping~$[F]$ and~$[G]$ as milestones of~$[H]$, we have to change the two others in
order to get the opposite direction along~$W$ with respect to that of~$V$: call this 
transformation of~$\Delta$ into this new pattern around~$[H]$ {\bf quasi-reflection}
of~$\Delta$ into~$[H]$.
 
\vskip 10pt
\vtop{
\vspace{-10pt}
\setbox110=\hbox{\epsfig{file=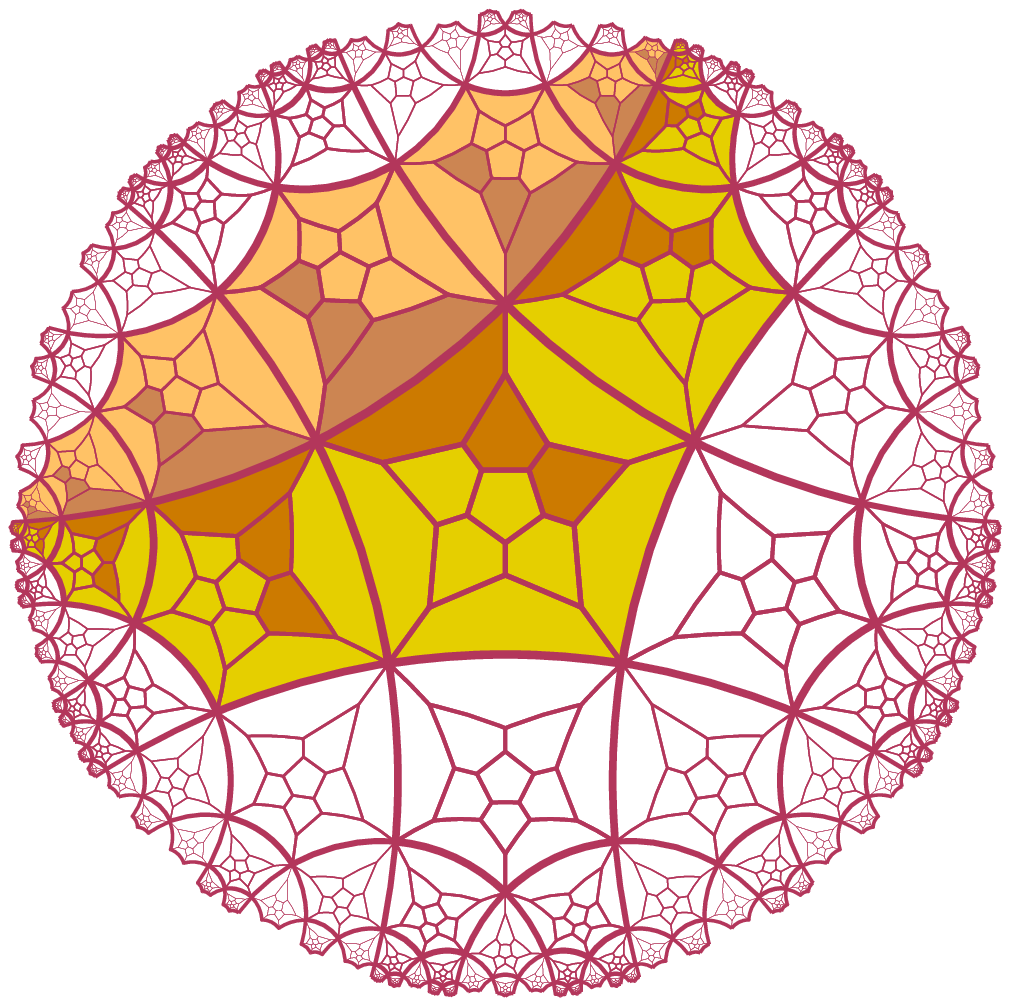,width=240pt}}
\setbox115=\hbox{\epsfig{file=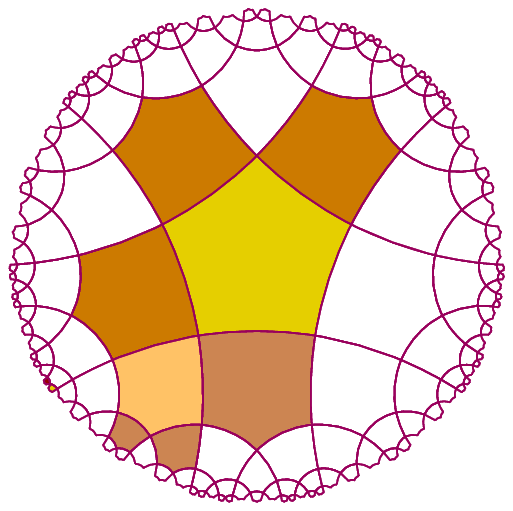,width=120pt}}
\setbox118=\hbox{\epsfig{file=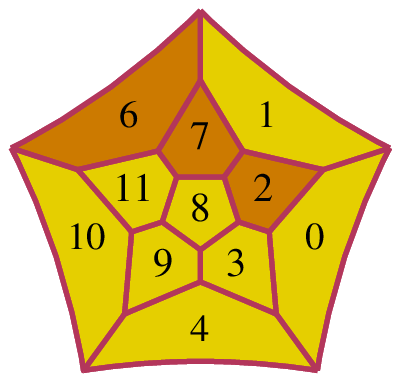,width=120pt}}
\ligne{\hfill
\PlacerEn {-175pt} {0pt} \box110
\PlacerEn {40pt} {-20pt} \box115
\PlacerEn {40pt} {100pt} \box118
\hfill
}
\vspace{-5pt}
\begin{fig}
\label{dodec_vert}
\leurre
Left-hand side:
The new vertical ways with two tracks.
In yellow, one direction; in brown, the opposite direction.
\vskip 2pt
\noindent
Right-hand side:
\vskip 0pt
up: the numbering of the faces in a dodecagrid of the tracks;
\vskip 0pt
down: a cut of the tracks in the plane of the face~$10$ of the central cell.
\end{fig}
}
\vskip 10pt
   To do that, number the faces of~$\Delta$ as indicated in the above picture of the
right-hand side of Figure~\ref{dodec_vert}. The face which is almost surrounded by the
milestones is always face~1 and the ballast is on the face~5 of~$\Delta$, as well as
of~$[H]$. Moreover, the numbers of the faces around face~11 are increasing while 
clockwise turning around face~11. Accordingly, the milestones of~$\Delta$ are on
faces~5, 2, 6~and~7. Taking the same conventions for~$[H]$ and taking
into account that $[H]$ is below~$\Pi_0$, we can see that the milestones of~$[H]$
are on its faces~0, 2, 5~and~7. Applying this transformation onto each dodecahedron
of~$W$, we get the track in the opposite direction which is illustrated by the left-hand 
side of Figure~\ref{dodec_vert}. In the figure, we imagine that~$\Pi_0$ is transparent,
so that $W$~can be seen through it. We also imagine that the milestones associated with
the faces~$G$ are also transparent: otherwise, the tiles of~$W$ could not be seen.
\vskip 10pt
   For the horizontal tracks, we apply the same idea, but its realization is more complex.
We already know from~\cite{mmarXiv3} that the implementation of a horizontal line is
more complex than that of a vertical line. Remember that there are two kinds of tiles
in a horizontal segment: we have straight elements, those which are present in a vertical
segment, and corners. In Figure~\ref{dodec_horiz}, one of the two tracks consist of
yellow and green tiles. The yellow tiles are the straight elements and the green ones are
the corners. The corner is represented by the right-hand side picture of 
Figure~\ref{dodec_horiz}. However, we have four kinds of corners. Some of
them have a milestone under face~0 and the others do not have it. Also in some of them,
face~9 has no milestone but face~10 does have while in the others, the situation is
opposite: there is a milestone on face~9 but there is none on face~10.
\vskip 10pt
\vtop{
\vspace{-10pt}
\setbox110=\hbox{\epsfig{file=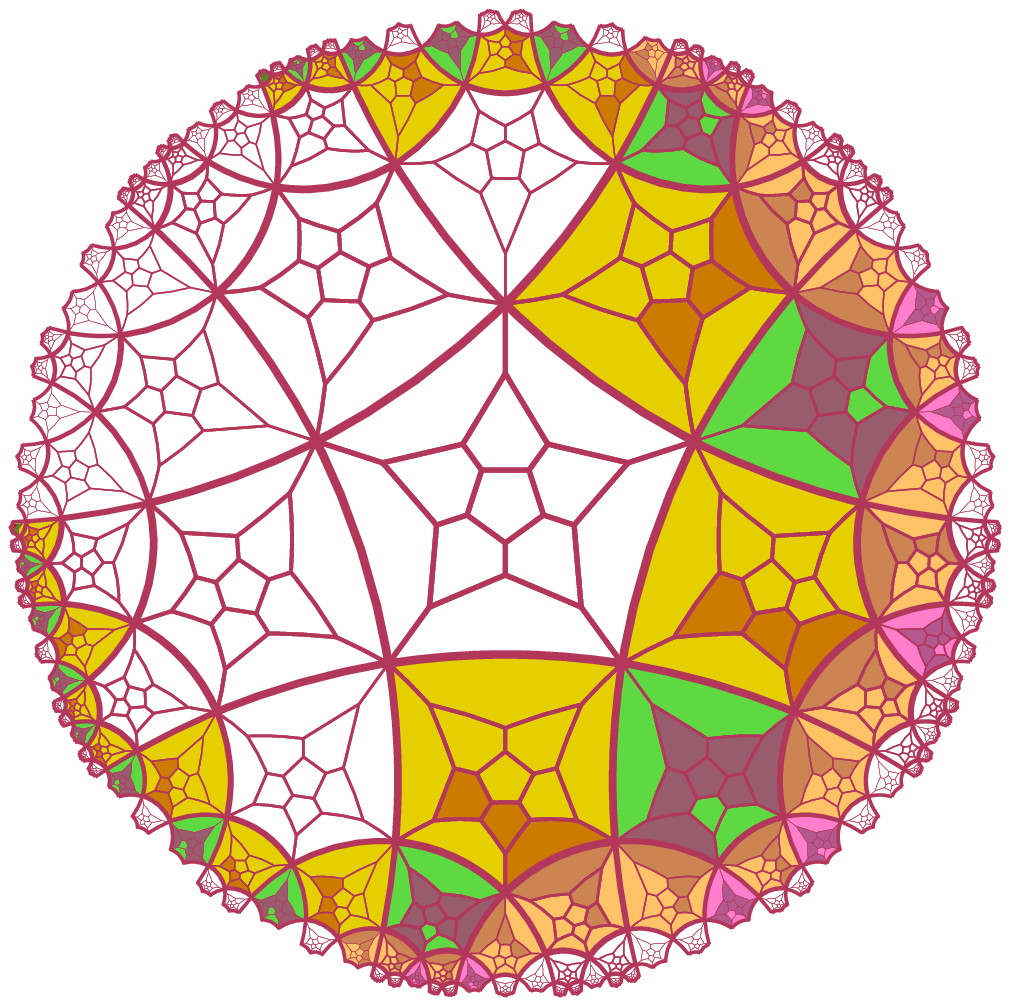,width=240pt}}
\setbox118=\hbox{\epsfig{file=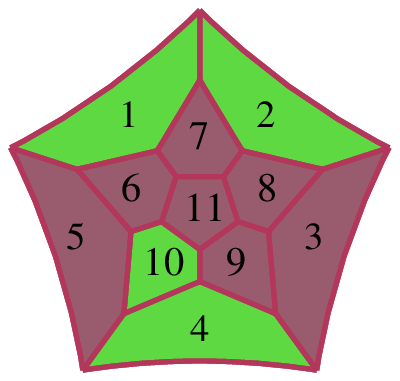,width=120pt}}
\ligne{\hfill
\PlacerEn {-175pt} {0pt} \box110
\PlacerEn {45pt} {100pt} \box118
\hfill
}
\vspace{-5pt}
\begin{fig}
\label{dodec_horiz}
\leurre
The new horizontal ways with two tracks.
In yellow and green, one direction. In brown and purple, the opposite direction.
Note that here we have straight elements and corners. We also have two kinds
of horizontal elements as well as two kinds of corners.
\end{fig}
}
\vskip 10pt
   We shall soon see the reasons of these distinctions.

   As for a vertical segment, the ballast of the return track of a horizontal segment 
consists of the milestones of the direct track which are in contact with the plane
of their ballast. Now, this requires to carefully define the corners and the
straight elements. 

   From the study of~\cite{mmarXiv3}, a horizontal segment, we shall later say a 
{\bf direct} horizontal segment, is a word of the form $(SQ)^a$, with $a$ a positive 
integer, where $Q$~is a corner and $S$~is a straight element. We say that $SQ$ is 
the {\bf basic unit}. In such a sequence of units, the quarter is connected to its 
neighbours in the chain by its faces~1 and~2, while the straight element is connected 
by its faces~1 and~4 or its faces~1 and~10. Whether it is~4 or~10 depends on the 
status of the face~5{} in the pentagrid defined on~$\Pi_0$ by the dodecagrid: it is 
face~4 when the face~5 is a white node, it is a face~10 when the face~5 is a black node. 
This distinction is relevant in the construction of the return track.

   First, consider the case when the face~10 of the straight element is in contact with
the corner of the basic unit. Denote by~$A$ the straight element and by~$B$
the corner.
As the face~10 of~$A$~is in contact with~$B$, consider the side~$s_6$ of face~6 which is
the single side of this face lying on~$\Pi_0$. The quasi reflection of~$A$ in~$s_6$
defines a dodecahedron~$A_6$ surrounded by milestones as in a straight element and the
face~1 of~$A_6$ defines a direction which is opposite to that defined by the face~1 of~$A$.
Accordingly, we consider~$A_6$ as belonging to the expected return track. Besides its 
face~5, no other face of~$A$ than its face~6 is in contact with~$\Pi_0$. Consider~$B$.
Outside its face~0, three of its faces are in contact with~$\Pi_0$: they are faces~3, 4
and~5. The intersections with~$\Pi_0$ are sides which we call~$s_3$, $s_4$ and~$s_5$ 
respectively. Let $B_3$, $B_4$ and $B_5$ the images of~$B$  in the quasi-reflection 
in~$s_3$, $s_4$ and~$s_5$ respectively. As the face~0 of~$B$ is shared by
a milestone, the same analysis as previously holds, and we can put around each $B_i$,
$i\in\{3,4,5\}$ four milestones. In the cases when $i$~is in $\{3,5\}$, the face~$i$
of~$B$ is shared by a milestone which we can define as the ballast of~$B_i$
and then the ballast of~$B$ is another milestone of~$B_i$. We define the two remaining
milestones of~$B_i$ thanks to the numbering which we already defined. The situation
for~$B_4$ is different: the face~4 of~$B$ is not shared by a milestone and so, the
ballast of~$B_4$ should not be a milestone. In order to keep the constraint which we
just noticed for~$B_4$, we define the face of~$B_4$ in contact with~$\Pi_0$ to be face~0.
Then we put milestones in the faces~2, 5, 6 and~7 of~$B_4$, its face~2 being in the plane
of the face~4 of~$B$. Now, we can see that $B_3$ and $B_4$ are not in contact with 
each other, and this is the same for~$B_4$ and~$B_5$. However, $B_5$~is in contact 
with~$A_6$: taking into account the construction of~$B_5$, the face~1 of~$A_6$ is in
contact with the face~3 of~$B_5$.

   Now, it is not difficult to see that the faces of~$B$, $B_5$ and~$B_4$ which lie 
on~$\Pi_0$ share a common point~$v_1$ ad that there is a fourth pentagon around~$v_1$
which is the reflection of the face~0 of~$B$ in~$v_1$. We can construct on this face
a dodecahedron~$Q_1$ which is in contact with both~$B_5$ and~$B_4$. We define it
as a corner which will be defined in a direction which is the same as those of~$B_5$
and~$B_4$. We may decide that the ballast of~$Q_1$ is not a milestone.

   Now, we can make a similar remark with $B$, $B_4$ and~$B_3$ whose faces on~$\Pi_0$
share a common point~$v_2$. We can construct another corner $Q_2$ in between~$B_3$
and~$B_4$ which is constructed in the same way as~$Q_1$.

   And so, in this situation, when the face~10 of~$A$ is in contact with~$B$,
we can see that this defines a part of the return track which consists of six dodecahedra:
four straight elements and two corners. Moreover, among the straight elements, three of
them have a ballast which is a milestone while this is not the case for the fourth one.

We can remark that the considered segment is a word $s_1s_1cs_0cs_1$, where $s_1$ defines a
straight element with a ballast which is a milestone and $s_0$ defines a straight
element where the ballast is not a milestone. We can also notice from the construction
that $_0$ is simply a rotated image of~$s_1$ around face~1.
  
   Second, consider the case when the face of~$A$ in contact with~$B$ is face~4. In this
case, besides face~5, $A$ has two faces in contact with~$\Pi_0$: faces~6 and~10. For~$B$,
the situation is the same as previously, with three faces, 3, 4 and~5{} in contact 
with~$\Pi_0$. Applying the quasi-reflection in the sides of the considered face lying
on~$\Pi_0$, this defines dodecahedra~$A_6$, $A_{10}$, $B_3$, $B_5$ and~$B_4$ which
can be defined as straight elements of the return track, having milestones as their
ballast which is in fact the milestone of the source-dodecahedron which lies on~$\Pi_0$,
above the plane. The numbering of the faces and the distribution of the milestones
are exactly as those described in the previous case for $A_6$, $B_5$ and~$B_3$. Now,
for~$A_{10}$ and~$B_4$, we have the same situation as $B_4$ in the previous case of
a basic unit: for these dodecahedra, their ballast should not be a milestone. Accordingly, 
for~$A_6$ and~$B_4$, we define the same numbering and the same distribution of the
milestones as for~$B_4$ in the previous case. Now, we can see that $A_6$ and~$A_{10}$
are not in contact and, similarly, as in the previous case, $B_3$ and $B_4$ as well
as $B_5$ and~$B_4$ are not in contact. We apply the same construction as previously,
yielding three corners these time, $Q_1$, $Q_2$ and~$Q_3$: $Q_1$ in between $B_4$ and~$B_5$,
$Q_2$ in between $B_4$ and~$B_3$ and $Q_3$ in between $A_6$ and~$A_{10}$, with exactly
the same characteristics as the corners which we obtained in the previous case.

   Accordingly, when the face~4 of~$A$ is in contact with~$B$, we can see that this defines
a segment of the return track which consists of eight dodecahedra: five straight segments
and three corners. Moreover, among the five straight elements, four of them have
a ballast which is a milestone while this is not the case for the fifth one. 

   As in the previous case, we can remark that the considered segment is a 
word $s_1cs_0s_1cs_0cs_1$, where $s_1$ and~$s_0$ as defined as previously.
  
   Now, it is plain that the face~1 of~$B_3$ is always in contact with the~$A_6$
defined by the next basic unit. Consequently, a direct segment of the
form $(SQ)^a$ defines a return segment of the form $(s_1\{cs_0\}s_1cs_0cs_1)^a$
where the occurrence of $cs_0$ is dictated by the number of the face of~$S$ which is
in contact with~$Q$.

   It is now time to deal with the motion of the particle.

\subsection{The motion of the particle}

   In order to do that, we consider the case of a vertical segment which is the easiest
one.

   We have to define what is the direction of the motion. From the previous analysis,
define the numbering of the faces as in the right-hand side picture of 
Figure~\ref{dodec_horiz}. For a straight element of a direct track, the milestones
are on faces 0, 5, 6 and~7. They are on faces 2, 5, 6 and~7 for a straight element of
a return track. This also holds for horizontal tracks. In a vertical track, direct 
or return, the face on~$\Pi_0$ is always face~5. In a horizontal track, it is also 
most often face~5 but, from time to time, it is face~0. We have described when this
happens in a precise way.

   From now on, we define that face~1 of a straight element is an {\bf exit} for the
particle. The {\bf entry} may be either face~4 or face~10. Of course, we might decide
to chose the opposite convention. However, we shall see that this solution makes the 
things a bit easier. In some occasions, as we shall see a bit later, the straight 
element may be rotated around face~1 so that face~0 lies on~$\Pi_0$. In this case, 
the entries may be either face~4 or face~3. Face~10 remains possible but it will 
be never used. At last, another rotation is also use in which face~2 lies 
on~$\Pi_0$. In this latter case, the entries are face~3 or~8. Face~8 will be 
mostly used.

   For a corner, the situation is simple. There are two possible tiles as
the pattern of the milestones around the cell of the track is not symmetric. And so,
there is a pattern as the one which is presented by the right-hand side picture of
Figure~\ref{dodec_horiz} and the symmetric one with respect to the reflection
which exchanges faces~9 and~10{} in the same figure, leaving faces~0, 4, 7 and~11 
globally invariant. The particle passes through faces~1 and~2. As the just indicated
two patterns are not a rotated image of each other, we can use one pattern for the
direction from face~1 to face~2 and the other for the direction from face~2 to face~1.
We decide that the milestone which is on one of the faces~9 or~10 is put on the side
of the entry. Accordingly, when face~9 is not covered by a milestone, the entry is
face~1 and the exit is face~2. When face~10 is not covererd by a milestone, face~1
is now the exit and face~2 is the entry.

   Before turning to the switches, we have to pay a new visit to the bridges which were 
introduced in~\cite{mmJCA3D} and which were adapted to the new definition of tracks 
in~\cite{mmarXiv3}.

\subsection{The bridges}

   In~\cite{mmarXiv3}, we have indicated how to perform the crossing of two vertical 
segments, as we may always assume that crossings can be performed in this way.
To avoid crossings, me decide that one track will pass in another plane of the $3D$~space
in order to avoid the other track and to go back to its way. As each segment consists of
two tracks, and although the tracks are very close to each other, we cannot represent a
whole bridge within a single figure.

   Consider two segments~$S_1$ and~$S_2$. We may assume that in both segments, 
the face~5 of the dodecahedra involved in both traks are all in the same plane~$\Pi_0$. 
Let $\ell_1$, $\ell_2$, be the guidelines of~$S_1$, $S_2$ respectively. Assume that
$\ell_1$~and~$\ell_2$ meet at some point~$P$. This point is a vertex of a pentagon
of~$\Pi_0$ as $\ell_1$ and~$\ell_2$ are two lines of the pentagrid defined on~$\Pi_0$ by
the dodecagrid. There are four pentagons around~$P$. From this, we can see that if
we keep one segment unchanged, the other has to avoid four dodecahedra of the other
segment: two of them above~$\P_0$ and the two others below. Let $S_2$~be unchanged 
and let us consider the change needed for~$S_1$. 

Let us introduce coordinates from~$Z\!\!\!Z$ on both tracks of~$S_1$. As one track is 
above~$\Pi_0$ and the other is below the plane, they will be denoted by~$T_u$ 
and~$T_\ell$ respectively. A dodecahedron of a track is thus numbered $T_u[n]$
or~$T_\ell[n]$. To fix things, we decide that $P$~is shared by $T_u[0]$, 
$T_u[$$-$$1]$, $T_\ell[0]$ and~$T_\ell[$$-$$1]$, the projection of $T_u[0]$
onto~$\Pi_0$ being a neighbour of the projection of~$T_\ell[$$-$$1]$ onto the same
plane. Now, we shall have two bridges~$b_u$ and~$b_\ell$: $b_u$ for the upper track 
will pass above~$\Pi_0$ and $b_\ell$~will pass below~$P_0$. In order to define the 
bridges, we consider the plane~$\Pi_1$ which contains~$\ell_1$ and which is perpendicular 
to~$\Pi_0$.

The dodecahedra of the tracks will mostly have their face~5 on~$\Pi_1$,
and the tracks will be in different half-spaces with respect to~$\Pi_1$.
The dodecahedra of~$b_u$ replace the dodecahedra $T_u[i]$ with $i\in[$$-$$3,2]$
and, similarly, the dodecahedra of~$b_\ell$ replace the dodecahdra $T_\ell[i]$,
with the same values for~$i$. 
The dodecahedra of~$b_u$ and~$b_l$ follow a kind of half-circle
in~$\Pi_1$ which is somehow truncated at the ends. Most of these dodecahedra are at 
a distance~6 or~7 from~$P$, $T_u[0]$,
$T_u[1]$, $T_\ell[0]$ and~$T_\ell[1]$ being at distance~0 by definition. A slight
modification allows us to reduce a bit the number of needed dodecahedra. In order to 
better see this point, Figures~\ref{end_projP0} and~\ref{end_projP1} represent the
ends of both~$b_u$ and~$b_\ell$. Figure~\ref{end_projP0} is a projection onto~$\Pi_0$
of the ends of~$b_u$ and~$b_\ell$: we can see there the last dodecahedron of the tracks
of~$S_1$. Figure~\ref{end_projP1} is a projection of these ends onto~$\Pi_1$. On this 
figure, we have represented four more dodecahedra of the upper bridge at each end in order
to better see the situation.

\vskip 10pt
\vtop{
\vspace{-10pt}
\setbox110=\hbox{\epsfig{file=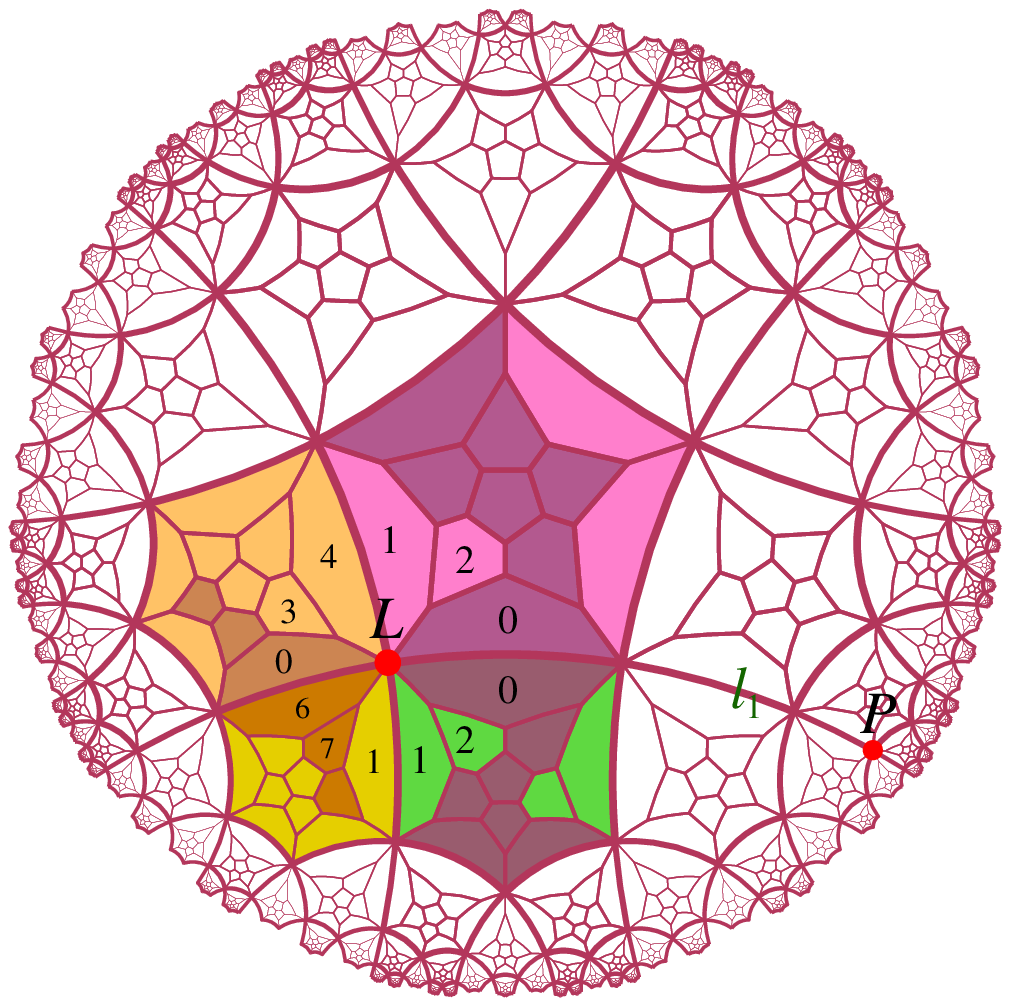,width=180pt}}
\setbox112=\hbox{\epsfig{file=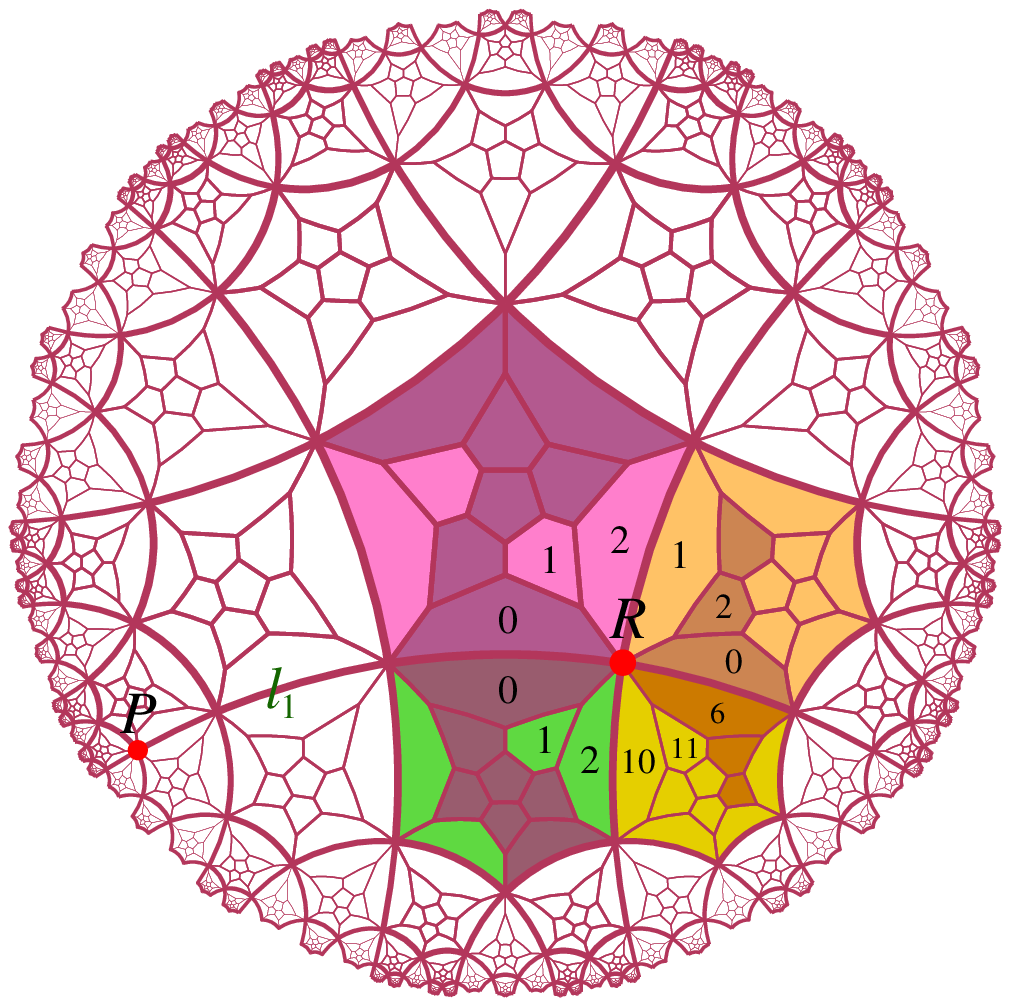,width=180pt}}
\ligne{\hfill
\PlacerEn {-180pt} {0pt} \box110
\PlacerEn {5pt} {0pt} \box112
\hfill
}
\vspace{-5pt}
\begin{fig}
\label{end_projP0}
\leurre
The ends of the bridges: projection onto~$\Pi_0$. The numbers given to the faces
around their common point are enough to find out all the numbers of the faces.
\end{fig}
}
\vskip 10pt
Denote by $b_u[1]$~the dodecahedron of the bridge which
replaces $T_u[$$-$$3]$ and similarly by $b_\ell[1]$ the one which replaces
$T_\ell[2]$. For the other end, denote by $b_u[$$-$$1]$ and $b_\ell[$$-$$1]$ the dodecahedra of~$b_u$ and $b_l$ respectively which replace $T_u[2]$ and~$T_\ell[$$-$$3]$ respectively.
This will allow us to number the dodecahedra of~$b_u$ and~$b_\ell$ which are close to their
ends.
Both faces on~$\Pi_0$ of these dodecahedra share a point~$L$ or~$R$ 
with their neighbours of~$T_u$ and~$T_\ell$ respectively. They are corners which are 
placed with face~5 on~$\Pi_0$ instead
of face~0. For both $b_u[1]$ and $b_\ell[1]$, face~0 is on~$\Pi_1$ so that these corners 
send the particle onto~$\Pi_1$ or they expect it from this plane. 

   Figures~\ref{end_projP0} and~\ref{end_projP1} describe the departure/arrival of 
each end of the bridges. Figure~\ref{end_projP0} shows the projections of~$b_u[1]$ and 
the corresponding dodecahedron of~$b_\ell$ onto~$\Pi_0$. Note that on this projection,
$b_u[1]$ is seen from above while $b_\ell[1]$ is seen from below, as if $\Pi_0$ would be
transparent. We have a similar phenomenon in Figure~\ref{end_projP1}. The tiles
of~$T_u$ and~$b_u$ which are projected onto $\Pi_1$ are projected from the front of~$\Pi_1$
with respect to the reader. The tiles of~$T_\ell$ and~$b_\ell$ are projected from behind
the plane. The consequence of these characteristics of the projections is that the numbering
of the faces is clockwise on the projections corresponding to~$T_u$ and~$b_u$. They are
counter-clockwise on the projections corresponding to~$T_\ell$ and~$b_\ell$.

\vtop{
\vspace{-10pt}
\setbox110=\hbox{\epsfig{file=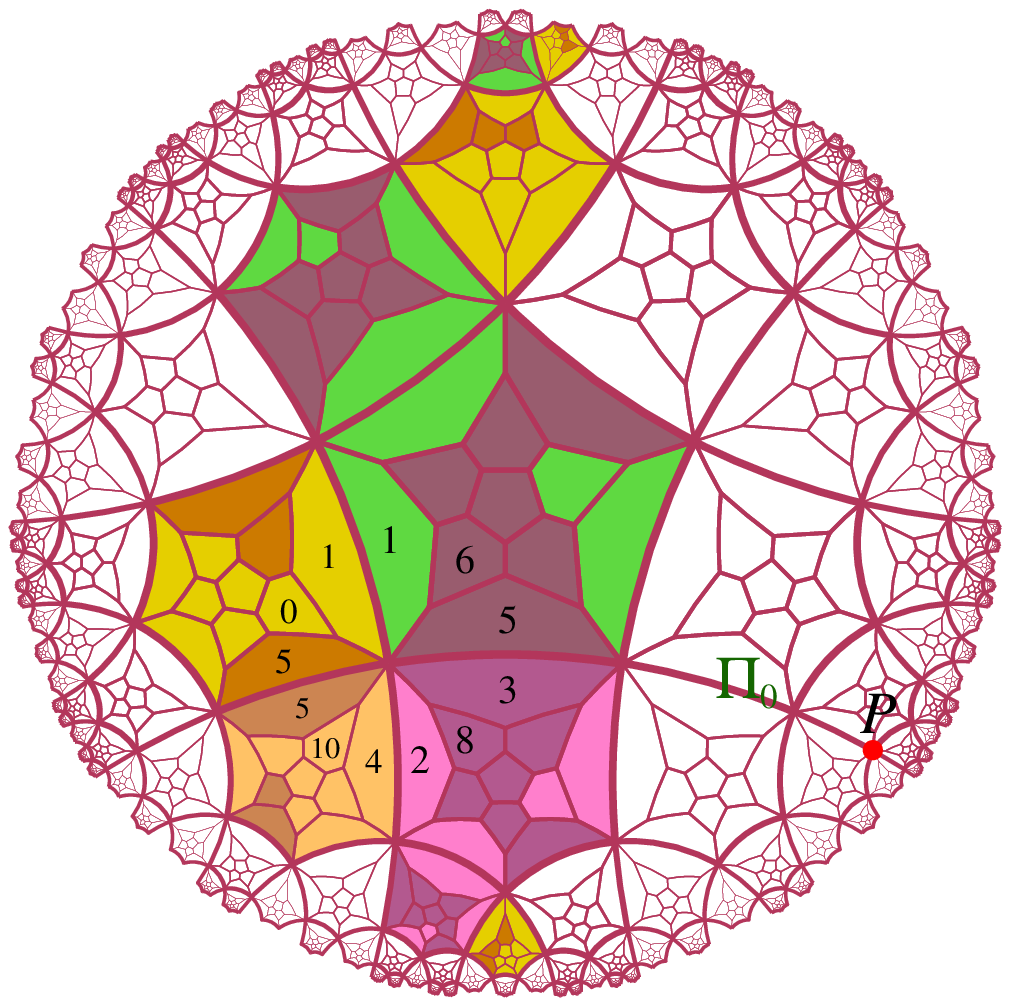,width=180pt}}
\setbox112=\hbox{\epsfig{file=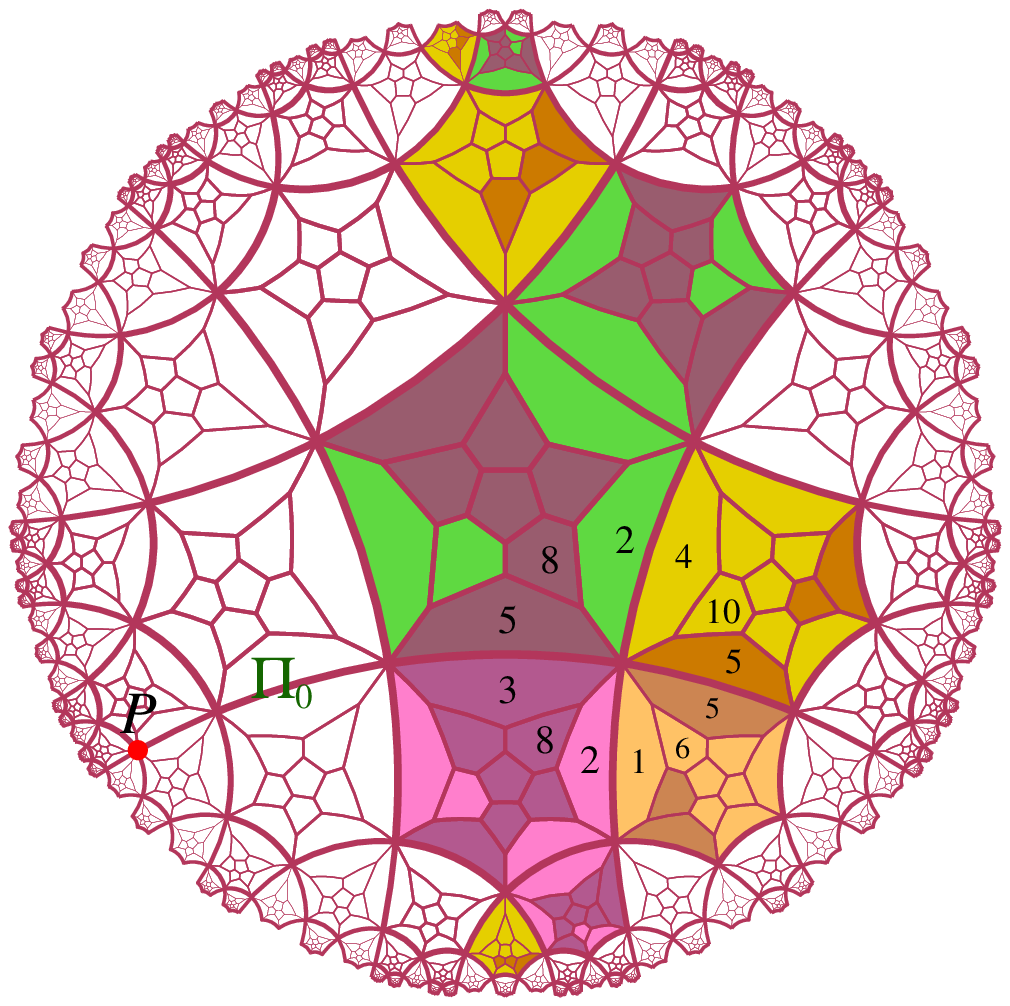,width=180pt}}
\ligne{\hfill
\PlacerEn {-180pt} {0pt} \box110
\PlacerEn {5pt} {0pt} \box112
\hfill
}
\vspace{-5pt}
\begin{fig}
\label{end_projP1}
\leurre
The ends of the bridges: projection onto~$\Pi_1$. The numbers given to the faces
around their common point are enough to find out all the numbers of the faces.
\end{fig}
}
\vskip 10pt
   It is not difficult to see that it is possible to join the two ends represented
by Figure~\ref{end_projP1}. Indeed, $b_u[5]$ and~$b_u[$$-$$5]$ are at distance~5 
from~$P$. Accordingly, they are on level~5 of two Fibonacci trees which are rooted
at the pentagons which share~$P$ in the half-plane of~$\Pi_0$ which contains~$b_u$.
The horizontal segment which we defined in this section follows such a level. In this case,
the yellow tiles are on level~5 while the green ones are on level~6. We also know
that the track going from~$b_u[5]$ to~$b_u[$$-$$5]$ along~$b_u$ is a word of the form
${\cal S}({\cal QS})^+$, where $\cal S$ represents a straight element and $\cal Q$
representes a corner. Accordingly this portion of half-circles completes the part of
the bridge which is not represented on the figure. By symmetry, we do the same 
for~$b_\ell$.

   We are now ready to study the implementation of the switches.
 
\section{The scenario of the simulation}
\label{scenario}

   We shall examine the three kinds of switch successively. We start with the trivial
case of the fixed switch. We go on with the rather easy case of the flip-flop switch
and we complete the study by the rather involved case of the memory switch.

\subsection{Fixed switches}
\label{fixed}

   As known from Section~\ref{railway}, we have a passive switch only to implement.
The definition of the motion of the particle by a passage from face~4 or~10 to face~1
shows that there is nothing to do but abutting the two arriving tracks to the exiting one.
This is illustrated by Figure~\ref{dodec_fix}. It is enough to make the central cell
symmetric, which is easy to realize: it is enough to rotate a straight element around its
face~1{} in such a way that the face lying on~$\Pi_0$ is now face~0. 
\vskip 10pt
\vtop{
\vspace{-10pt}
\setbox110=\hbox{\epsfig{file=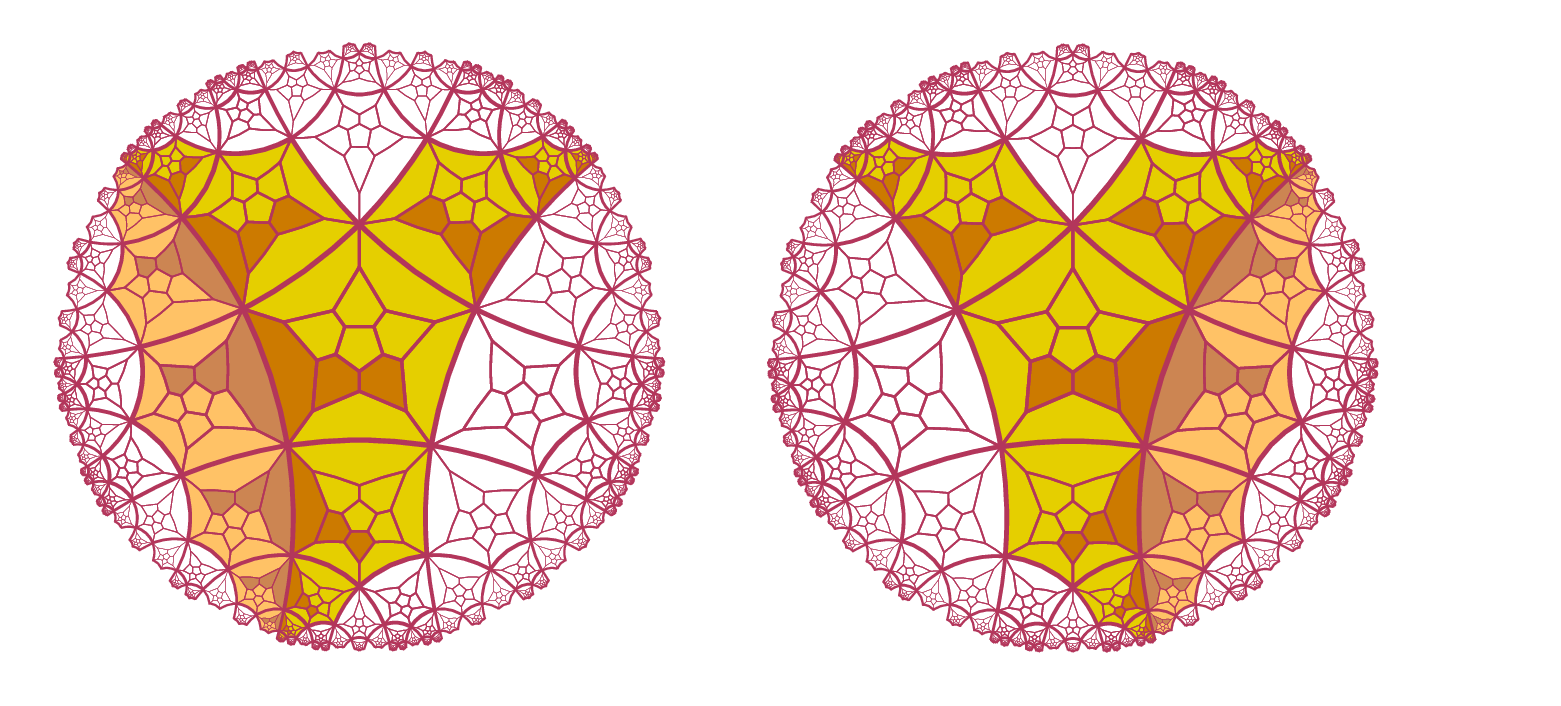,width=360pt}}
\ligne{\hfill
\PlacerEn {-163pt} {0pt} \box110
\hfill
}
\vspace{-5pt}
\begin{fig}
\label{dodec_fix}
\leurre
Two fixed switch:  left-hand side and right-hand side. Note the symmetry of the figure.
Also note that this requires straight elements only.
\end{fig}
}
\vskip 10pt
  Now, it is easy to put the return track, either along one arriving track or along the 
other: Figure~\ref{dodec_fix} shows that the orientation of the milestones presented in 
the figure makes both constructions possible. The changes needed to transform one picture
of the figure into the other are easy to perform. Note that we have here an easy realization
of two versions of a fixed switch while in many previous papers we had only a right-hand
side fixed switch, the other fixed switch being simulated by the right-hand side one 
followed by a crossing.

   We shall see that this easy implementation will help us to realize a rather simple
implementation of the passive memory switch.

\subsection{Flip-flop switches}
\label{flip-flop}

   Now, we have to realize the flip-flop switch. We can easily see that it is not
enough to reverse the straight elements with respect to the previous figure in order
to realize the switch. This has to be done for the tracks but it does not work for
the central cell, the cell which the three tracks abut. The central cell must have a 
specific pattern. We decide to append just one additional milestone and we change a bit the
pattern, making it symmetric and significantly different from that of the fixed switch.

   This fixes the frame but now, we have to implement the mechanism which first,
forces the particle to go to one side and not to the other one and then, to change the
selected track. 
\vskip 10pt
\vtop{
\vspace{-10pt}
\setbox110=\hbox{\epsfig{file=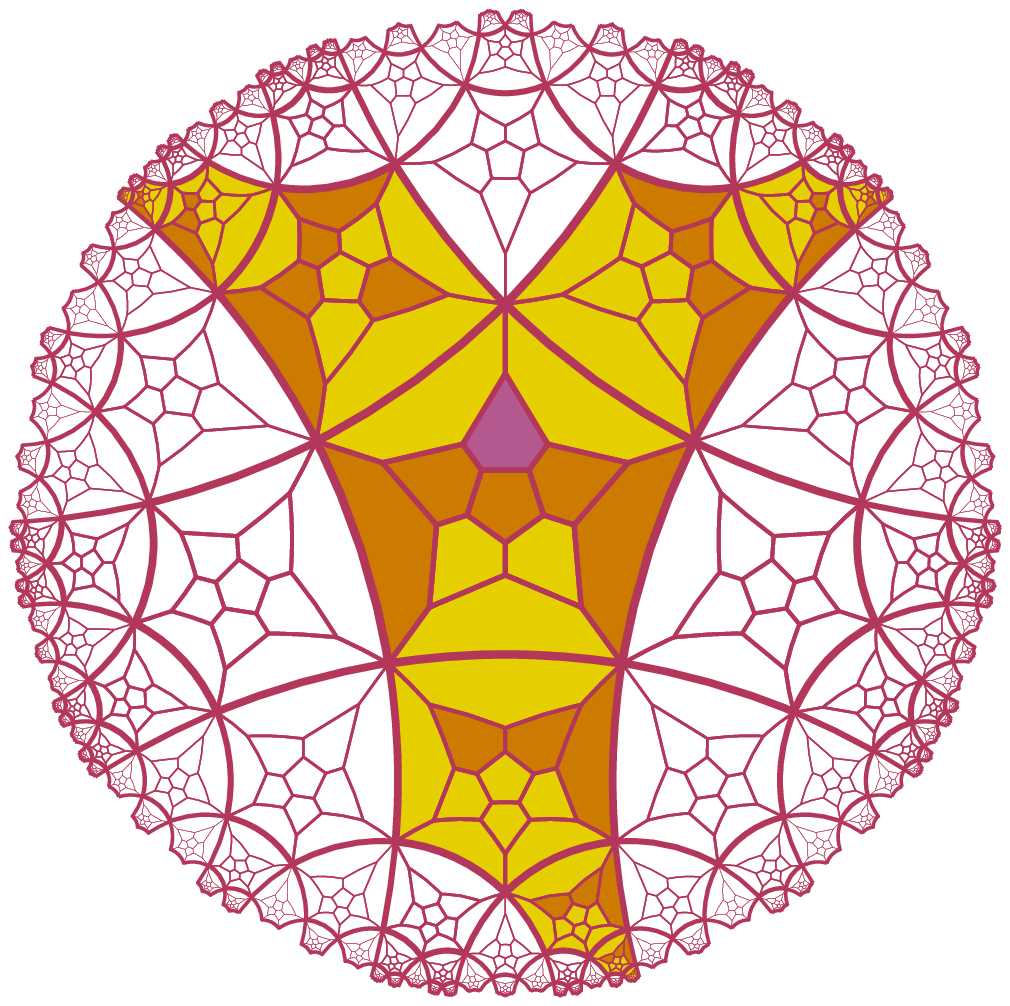,width=240pt}}
\ligne{\hfill
\PlacerEn {-125pt} {0pt} \box110
\hfill
}
\vspace{-5pt}
\begin{fig}
\label{dodec_flip_flop}
\leurre
The flip-flop switch: here the selected track is the right-hand side one.
Note that three cells have a particular patter: the central cell and the two tracks
marking the abutting of the ways~$a$ and~$b$.
\end{fig}
}
\vskip 10pt
   Let us remember notations and let us define new ones. We remember that three
tracks abut a switch. In section~\ref{railway}, we called them~$u$, $a$ and~$b$,
where, in an active passage, $u$ is before the switch and $a$ with~$b$ are together
after the switch, $a$ on the right-hand side, $b$ on the left-hand one. Also number 
the cells in the following way, followed by the computer program: 1, 2 and~3 are the 
cells of~$u$, 3~being the cell which abuts the central cell which receives number~4. 
Next, 5, 6 and~7 are the cells of~$b$ and 8, 9 and~10 are those of~$a$, with
5 and~8 abutting~4. Cell~5 is called the {\bf entry} of~$b$ and cellé8 is called that
of~$a$. At last, we distinguish three other cells called 11, 12 and~13.
To locate them, we number the faces of~4, 5 and~8. In all these cells, face~1 is the yellow
outer face whose both neighbouring faces inside the same pentagone are brown. Then, the 
other faces are increasingly numbered by clockwise turning around the cell, first the
outside ring of faces and then the inside one, completing with the inmost face.
Considering this numbering, cells~11, 12 and~13 are the cells which are put on the face~9
of the cells~5, 8 and~4 respectively. Cells~11 and~12 are the sensors of cells~5 and~8 
respectively.

   Now, cells~11 and~12 signalize the selected track: one of them is white and the
other is black. The selected track is the one for which the sensor of its entry is
white. Cell~13 is called the {\bf controller}. It is usually white and when it
flashes to black, cells~11 and~12 simulatneously change their state to the opposite one.
Cell~13 is black just for this action: after that it goes back to the white state.

   This scenario is illustrated by Figure~\ref{part_flip_flop} as well as by the
trace of execution given by Table~\ref{t_flip_flop}.

\vskip 10pt
\vtop{
\begin{tab}\label{t_flip_flop}
\leurre
Trace of execution performed by the computer program to check the crossing of
a right-hand side flip-flop switch by the particle.
\end{tab}
\vspace{-12pt}
\grostrait
\vskip3pt
{\obeylines
\obeyspaces\global\let =\ \tt\parskip=-2pt
active crossing of a right-hand side flip-flop switch :
\vskip3pt
          1  2  3  4  5  6  7  8  9 10 11 12 13
\vskip3pt
time 0 :  W  B  W  W  W  W  W  W  W  W  B  W  W
time 1 :  W  W  B  W  W  W  W  W  W  W  B  W  W
time 2 :  W  W  W  B  W  W  W  W  W  W  B  W  W
time 3 :  W  W  W  W  W  W  W  B  W  W  B  W  B
time 4 :  W  W  W  W  W  W  W  W  B  W  W  B  W
time 5 :  W  W  W  W  W  W  W  W  W  B  W  B  W
}
\demitrait
}
\vskip 10pt
   How can cell~13 detect the situation? In fact, cells~11, 12 and~13 are not fixed 
cells. So, to change states according to the indicated scenario, they are themselves 
decorated by an appropriate pattern. It is the same pattern for cells~11 and~12 as
they play the same role, it is another pattern for cell~13. The patterns are illustrated 
by Figure~\ref{control_flip_flop}.

\vskip 10pt
\vtop{
\vspace{-10pt}
\setbox110=\hbox{\epsfig{file=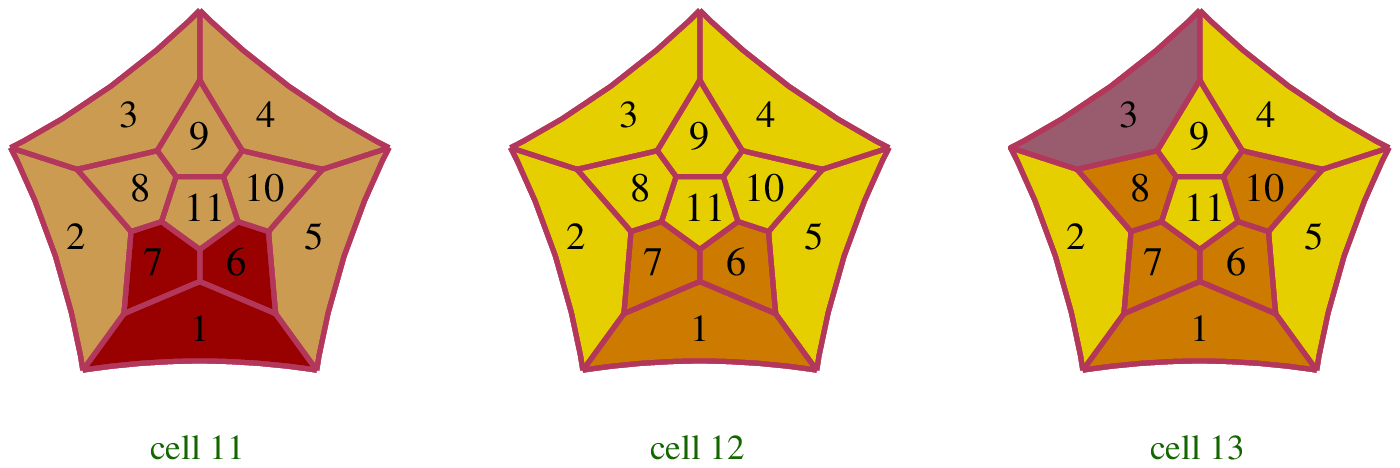,width=240pt}}
\ligne{\hfill
\PlacerEn {-125pt} {0pt} \box110
\hfill
}
\vspace{-5pt}
\begin{fig}
\label{control_flip_flop}
\leurre
The cells which control the working of a flip-flop switch. Cells~$11$, $12$ and~$13$
are placed on the face~$9$ of cells~$5$, $8$ and~$4$ respectively. Cells~$11$ and~$12$ are
mere signals. Cell~$13$ is a control unit. When it flashes, cells~$11$ and~$12$
change their states. Here, cell~$11$ is represented with darker colours in order to
indicate that cell~$11$ is black while cell~$12$ is black.
\end{fig}
}
\vskip 10pt
   In order to understand the working of theses cells, we have to take into
cnsideration that cells~4, 5 and~8 share a common vertex, that cell~4 and~5 share a
common face as well as cells~4 and~8. More precisely, the face~4 of cell~5 and
the face~3 of cell~4 are the same. Similarly, the face~3 of cell~8 and the face~4
of cell~4 are the same. We shall say that cell~4 can see cells~5 and~8 through its faces~3
and~4 respectively and that cells~5 and~8 can see cell~4 through their face~4 and~3 
respectively. As a consequence, as cells~11, 12 and~13 share also a common vertex due to
their position on cells~5, 8 and~4 respectively, we have that cell~13 can see cells~11
and~12 through its faces~3 and~4 respectively and that cells~11 and~12 can see cell~13 
from their faces~4 and~3 respectively. This property is represented in 
Figure~\ref{control_flip_flop} where in cell~13, face~3 has another colour: this simply 
indicates that the neighbour of this cell is black and the neighbour is cell~11. This 
holds for a flip-flop where the right-hand side track is selected. When the left-hand 
side track is selected, cell~11 is white, cell~12 is black and, accordingly, in cell~13, 
face~3 has a white neighbour and face~4 has a black one. 

   The trace of execution given in Table~\ref{t_flip_flop} shows that the depicted
scenario takes place actually. Figure~\ref{part_flip_flop} illustrates the motion
of the particle. It represents the action of the controller by materializing its
different states as additional states. This is to underline the fact that although
cell~13 is by itself either black or white, the presence of its pattern as shown
by Figure~\ref{control_flip_flop} make it possible to view these states as two new
additional states.

   Of course, we have the symmetric transformation when the particle crosses
a flip-flop where the selected track is the left-hand side one.
\vskip 10pt
\vtop{
\vspace{-10pt}
\setbox110=\hbox{\epsfig{file=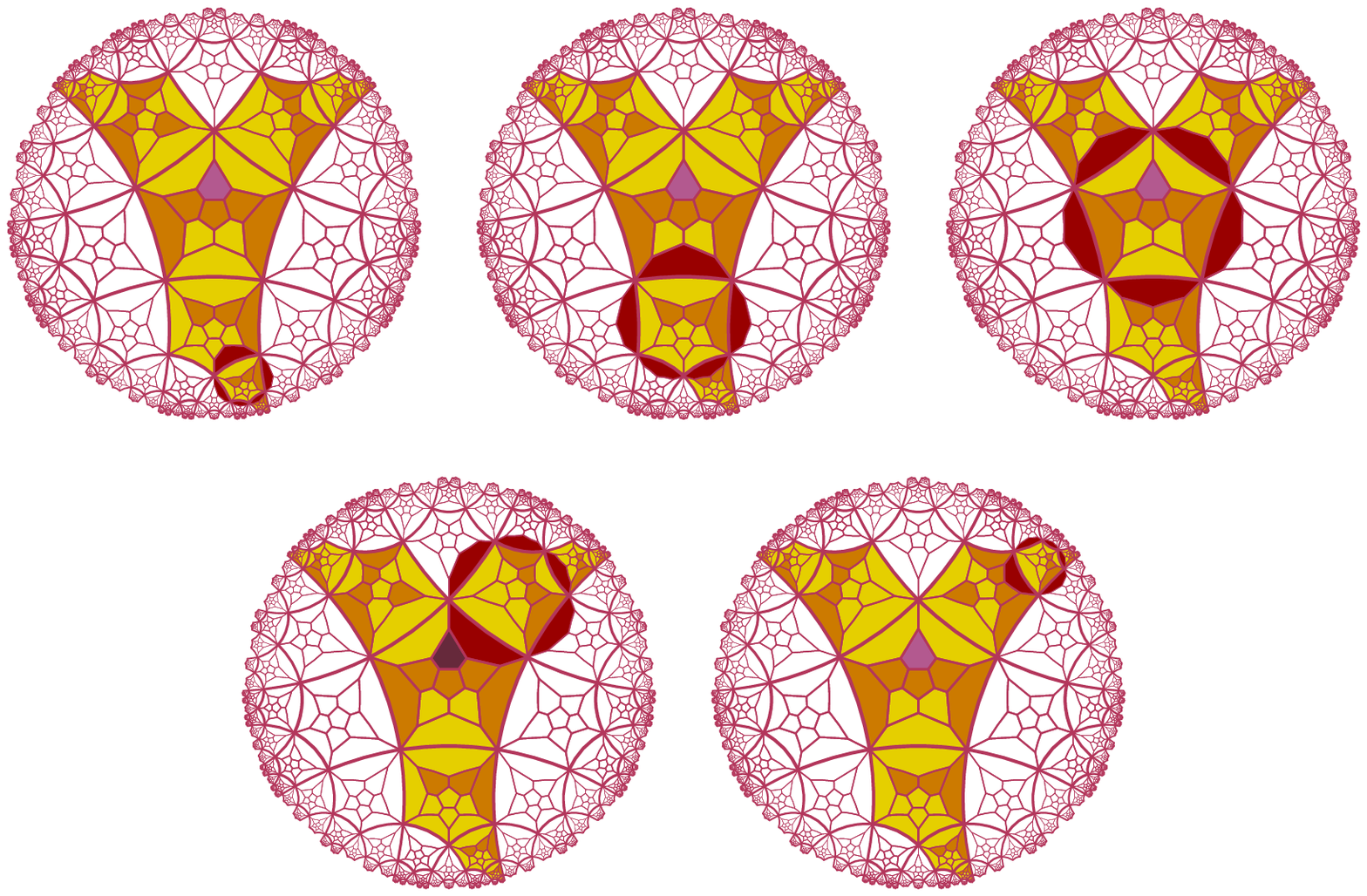,width=240pt}}
\ligne{\hfill
\PlacerEn {-125pt} {0pt} \box110
\hfill
}
\vspace{-5pt}
\begin{fig}
\label{part_flip_flop}
\leurre
Passage of the particle across a flip-flop switch. Note the changes of states
on the fourth and fifth pictures.
\end{fig}
}
\vskip 10pt

\subsection{Memory switches}
\label{memory}

   Now, we arrive to the most complex structure in our implementation. This is not due
to the fact that we have to impement two single-track switches: a passive one and an 
active one. It is mainly because these two single-track switches must be connected.
This comes from the definition of the memory switch: the selected track of the
active switch is defined by the last crossing of the passive switch.

   The name of the passive switch is a bit misleading. As will be seen, although
discreet, the passive switch has a definite action. If the particle happens to cross it
through the track which corresponds to the selected one in the active switch, it sends 
a signal which triggers the change of the selected track in the active switch. 
This can be performed thanks to a careful study of the previous switches to which we
now turn.

   Figure~\ref{dodec_memog} represents the passive memory switch. Here, the selected track
is the left-hand side one.
    
\vskip 10pt
\vtop{
\vspace{-10pt}
\setbox110=\hbox{\epsfig{file=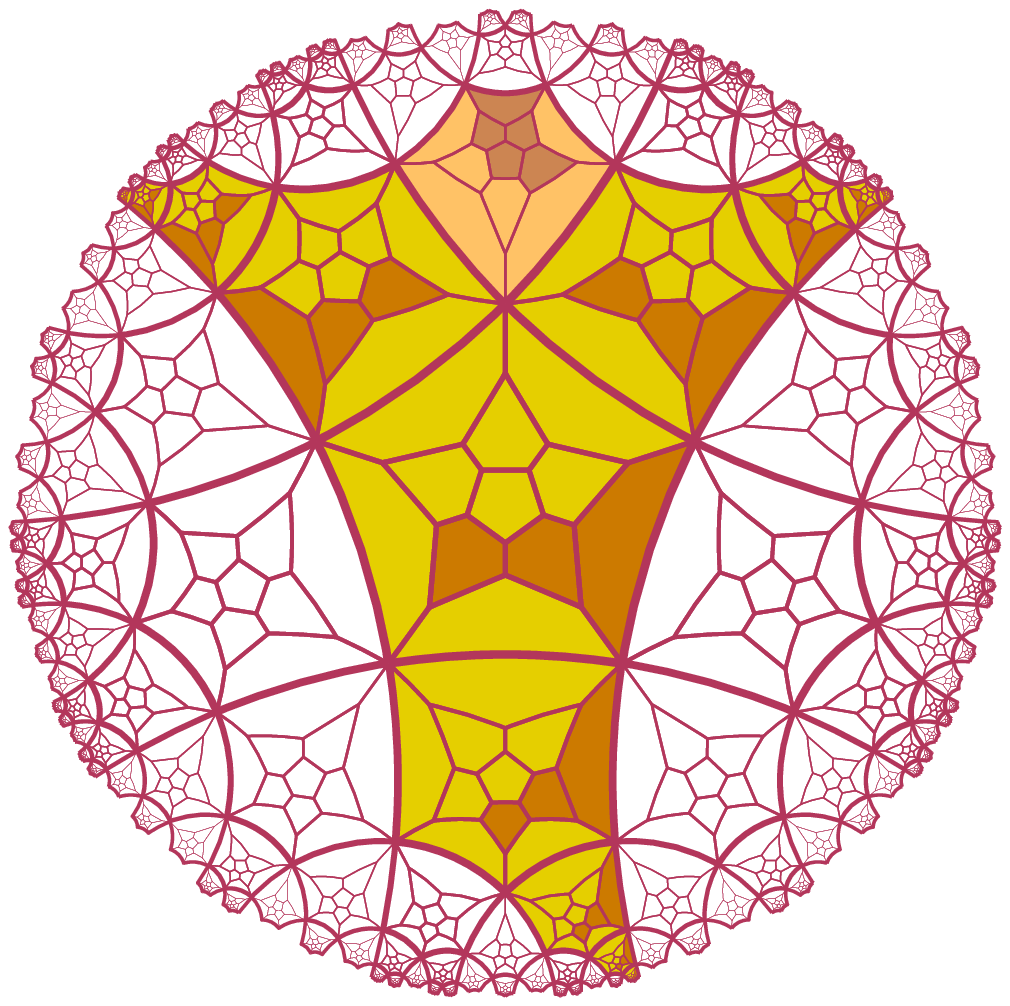,width=240pt}}
\ligne{\hfill
\PlacerEn {-125pt} {0pt} \box110
\hfill
}
\vspace{-5pt}
\begin{fig}
\label{dodec_memog}
\leurre
The passive memory switch. In this figure, the selected track is the 
left-hand side one. Nothing happens if the particle comes through this
track. If it comes through the other one, this will trigger the change of
selection, both in the passive and in the active switch.
\end{fig}
}
\vskip 10pt
   As in the case of the flip-flop switch, we number the cells involved in the passive
switch by taking the number they received in the computer program.

   This time, the cells from 1~to~3 are the cells of~$u$, 4~is again the central cell,
the cells from~5 to~7 are~$b$ and those from~8 to~11 are~$a$. Cell~4 can see cells~5 and~8
through its faces~3 and~4 respectively and, conversely, cells~5 and~8 can see cell~4 through
their face~4 and~3 respectively. Cells~11, 12 and~13 are very different from the cells
with the same numbers in a flip-flop switch. Here, cell~13 can be characterized as 
follows: let~$A$ be the common vertex of cells~4, 5 and~8. Above~$\Pi_0$, there are four 
dodecahedra sharing~$A$. We just mentioned three of them. The fourth one is cell~13, 
which is obtained from cell~4 by reflection in the plane orthogonal to~$\Pi_0$ which
passes through~$A$ and which cuts the faces~1 of cell~5 and~8 perpedicularly. Now, we 
can number the faces of cell~13{} in such a way that cell~13 can see cells~5 and~8 through 
its faces~4 and~3. Cells~11 and~12 are put on cell~13, on its faces~10 and~8. Cells~11
and~12 are the reflection of cells~5 and~8{} respectively in the edge which is shared 
by both the concerned dodecahedra. If the selected track is~$b$, cell~12 is black and 
cell~11 is white. If the selected track is~$a$, cell~11 is black and cell~12 is white.
Now we can describe what happens more clearly.
\vskip 10pt
\vtop{
\vspace{-20pt}
\begin{tab}\label{t_memogp}
\leurre
Trace of execution performed by the computer program to check the crossing of
a left-hand side passive memory switch by the particle. Here, the particle runs over the
non-selected track.
\end{tab}
\vspace{-12pt}
\grostrait
\vskip3pt
{\obeylines
\obeyspaces\global\let =\ \tt\parskip=-2pt
passive crossing of a memory switch, left-hand side,
  through the NON selected track :
\vskip3pt
          1  2  3  4  5  6  7  8  9 10 11 12 13 14 15
\vskip3pt
time 0 :  W  W  W  W  W  B  W  W  W  W  W  B  W  W  W
time 1 :  W  W  W  W  B  W  W  W  W  W  W  B  W  W  W
time 2 :  W  W  W  B  W  W  W  W  W  W  W  B  B  W  W
time 3 :  W  W  B  W  W  W  W  W  W  W  B  W  W  B  W
time 4 :  W  B  W  W  W  W  W  W  W  W  B  W  W  W  B
time 5 :  B  W  W  W  W  W  W  W  W  W  B  W  W  W  W
time 6 :  W  W  W  W  W  W  W  W  W  W  B  W  W  W  W
time 7 :  W  W  W  W  W  W  W  W  W  W  B  W  W  W  W
}
\demitrait
}
\vskip 10pt
\vskip 10pt
\vtop{
\vspace{-10pt}
\setbox110=\hbox{\epsfig{file=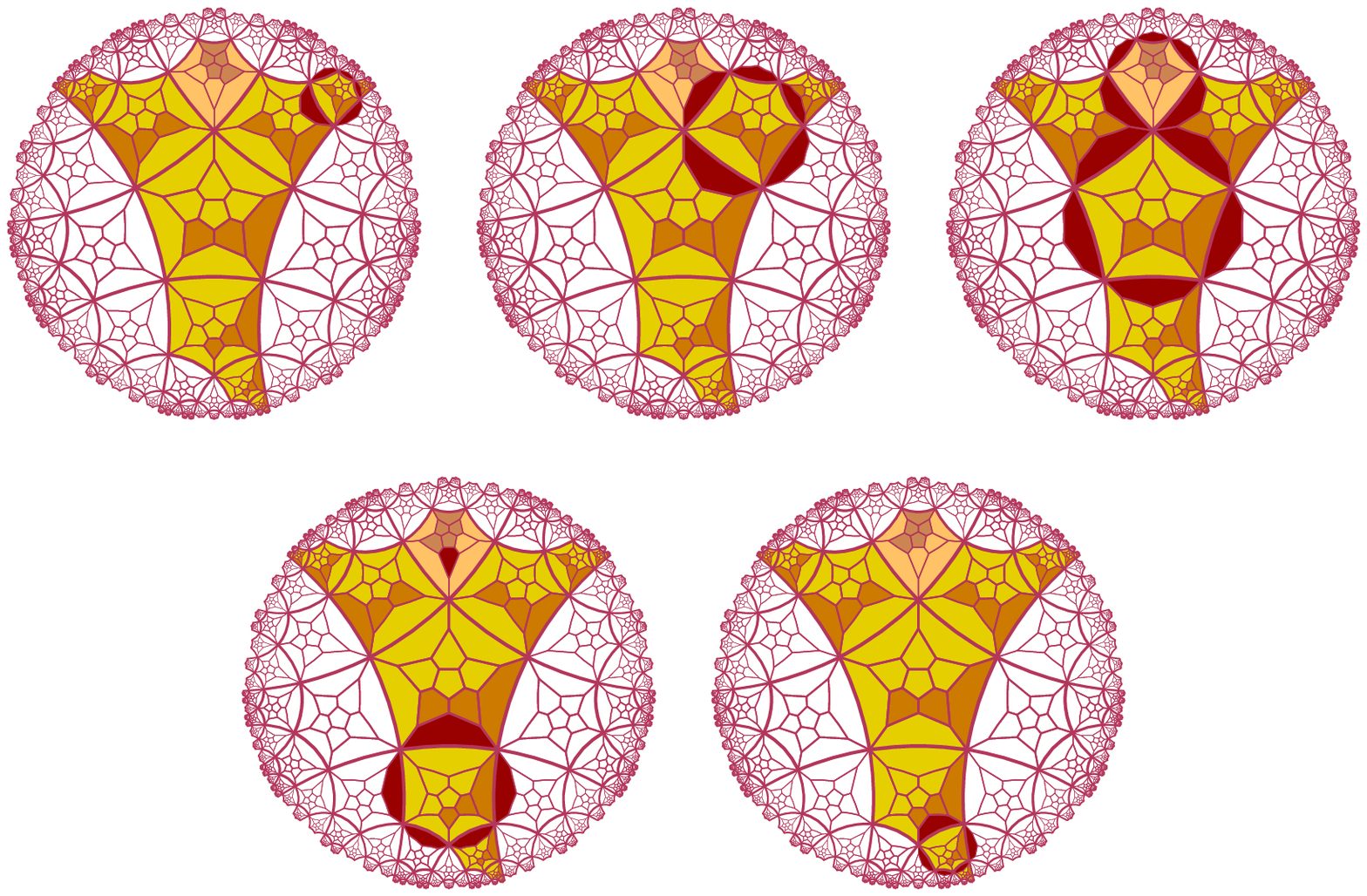,width=240pt}}
\ligne{\hfill
\PlacerEn {-125pt} {0pt} \box110
\hfill
}
\vspace{-5pt}
\begin{fig}
\label{part_memogp}
\leurre
Passage of the particle across a passive memory switch. Note the changes of states
on the third and fourth pictures.
\end{fig}
}
\vskip 10pt
   As required by the working of a memory switch, if the particle crosses the
passive memory switch through the selected track, nothing happens.

   And so, consider the case when the particle crosses the passive switch through the
non-selected track. When the particle is in~cell~8, cell~13 can see the particle and, as
its cell~12 is black, it knows that it must flash. This means that it becomes black 
for the next time and then returns to the white state at the following time. This can 
be seen in the trace of execution displayed by Table~\ref{t_memogp} as well as in
Figure~\ref{part_memogp}.
   When cell~13 flashes, cells~11 and~12 exchange their states: if it was black it becomes
white and conversely. 

But this is not the single change. The trace of Table~\ref{t_memogp}
mentions two additional cells: 14 and~15. They are the first two cells of a path
which conveys the flash emitted by cell~13 to the active switch in order that the
selected track of the active switch should also be changed. Cell~14 is put on cell~13,
on the face~9 of this cell precisely. This can be seen on the fourth picture of
Figure~\ref{part_memogp}: we can see that this cell becomes black just after the
flash of cell~13.

   This path introduces a new feature which does not exist in the simulations of the 
previous papers. As long as the flash does not reach the active switch, we have
two particles in the circuit: the one which plays the role of the locomotive introduced
in Section~\ref{railway}, and the one which is emitted by the sensor of the passive
memory switch. This new particle can be seen as a temporary copy of the first one.

   Let us now describe the path. But, to better understand this point, we have
to indicate how the active memory switch is implemented. In fact, it is implemented
as a flip-flop switch and, accordingly, cells are numbered in the same way as for
the flip-flop switch and cells~11, 12 and~13 are in the same connections with the
other cells. However, in the active memory switch cell~13 is very different from
the cell~13 of the fip-flop switch. This is materialized by the pattern of milestones
around the cell which is very different, see Figure~\ref{controllers}.  
\vskip 10pt
\vtop{
\vspace{-10pt}
\setbox110=\hbox{\epsfig{file=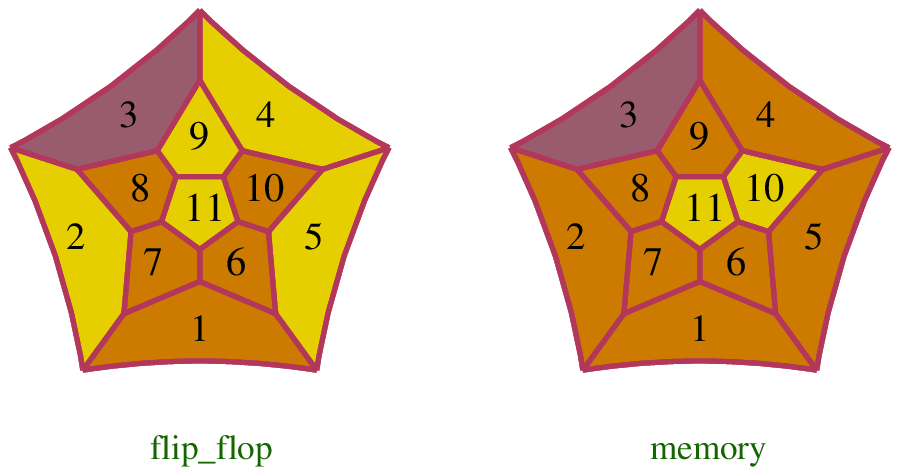,width=240pt}}
\ligne{\hfill
\PlacerEn {-125pt} {0pt} \box110
\hfill
}
\vspace{-5pt}
\begin{fig}
\label{controllers}
\leurre
Comparison between the controller of a flip-flop switch, left-hand side, and
the controller of an active memory switch, right-hand side.
\end{fig}
}
\vskip 10pt
   If we consider cells~1 to~10{} in a passive memory switch and in an active one,
up to a possible change of some numbers, their projection on~$\Pi_0$ is the same:
this is clear from the previous descriptions and from Figures~\ref{dodec_flip_flop}
and~\ref{dodec_memog}. Call this projection the {\bf basis} of the memory switch.
Consider the point~$P$ which is the common point of cells~4, 5 and~8. It is also
shared by cell~13{} in the passive memeory switch. Call~$\Delta$ the line which is
perpendicular to~$\Pi_0$ and which passes through the point~$A$ of~$\Pi_0$ 
which is shared by the face~4 and~5 of cell~4. It is a line of the dodecagrid:
it consists of edges of dodecahedras belonging to the tiling. 

We fix two points $A_p$ and $A_a$ on~$\Delta$ such that $A_p$ and~$A_a$ are vertices
of dodecahedra of the tiling and that $A$~is the mid-point of $[A_aA_p]$. We shall 
consider that $A_p$ is above~$A_a$. Now, we consider the planes~$\Pi_0^p$ 
and~$\Pi_0^a$ which are perpendicular to~$\Delta$ and which pass through~$A_p$ 
and~$A_a$ respectively. By our definitions, $A_p$~is in the half-space which is 
above~$\Pi_0^a$. We put a copy of the active memory switch onto~$\Pi_0^a$. This 
means that the face~0 of cell~4 is on~$\Pi_0^a$ and that cell~4 is in the half-space 
which is above~$\Pi_0^a$. Consider the planes $\Pi_4$ and~$\Pi_5$ which are 
perpendicular to~$\Pi_0^a$ and which contain the faces~4 and~5 of cell~4 
respectively. The intersection of these planes is the line~$\Delta$. Remember 
that $\Pi_0^p$ and~$\Pi_0^a$ are not parallel: as they have a common perpendicular 
they have no point in common, neither in the space nor at infinity. Next, we put 
a copy of the passive memory switch onto~$\Pi_0^p$ but in such a way that the 
cell~4 of the passive switch is in the half-space which is below~$\Pi_0^p$:
in this way, the cells~4 of both switches are contained in the intersection of
two half-spaces defined by~$\Pi_0^a$ and~$\Pi_0^p$. Also, we put the cell in 
such a way that its face~2 is on~$\Pi_5$ so that its face~3 is on~$\Pi_4$.
\vskip 10pt
\vtop{
\vspace{-10pt}
\setbox110=\hbox{\epsfig{file=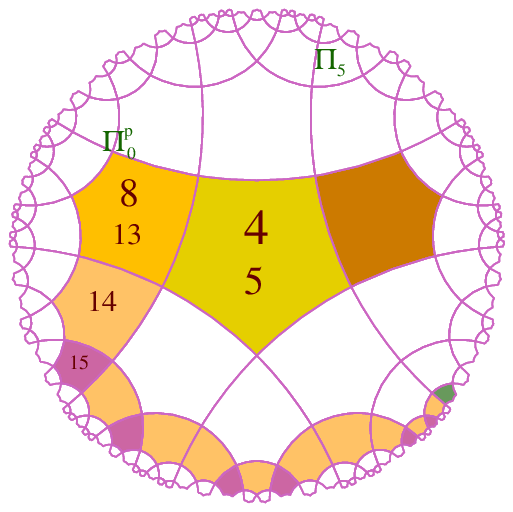,width=240pt}}
\ligne{\hfill
\PlacerEn {-125pt} {0pt} \box110
\hfill
}
\vspace{-5pt}
\begin{fig}
\label{memog_up_p4}
\leurre
Projection onto~$\Pi_4$ of a part of the path for the signal from the passive memory 
switch to the active one. We have put the number of the cells of the passive memory 
switch which are in contact with~$\Pi_4$. Note the change of levels of the path as it 
goes nearer~$\Delta$ which is the trace of~$\Pi_5$ on~$\Pi_4$.
\end{fig}
}
\vskip 10pt
   Figure~\ref{memog_up_p4} is a projection of the passive memory switch on~$\Pi_4$.  
The traces of~$\Pi_5$ and~$\Pi_0^p$ are indicated in the figure. Only four cells 
have a face on~$\Pi_4$: cells~4, 5, 8 and~13. This is why these numbers only 
appear in the figure for cells of the memory switch. Now, the common face of 
cells~4 and~5 is on~$\Pi_4$ and it is the same for the common face of cells~5 
and~13: this explains why the numbers of the cells sharing a face appear in the 
same yellow pentagons in the figure. We cannot see the other faces of the switch 
as they are not in contact with~$\Pi_4$. Cells~14 and~15 are the first cells of 
the path from the passive switch to the active one.
We can see them in Figure~\ref{dodec_memog}. Most cells of the path of the figure 
are in orange and in purple: this is to indicate that there are behind~$\Pi_4$ if 
the memory switch is mainly in front of the plane.  However, the path has to arrive 
to the dodecahedron which is on the face~9 of cell~4 of the active memory switch. 
Accordingly, at some point, the path must cross the plane~$\Pi_4$ in order to be 
in front of it. This is performed when the path reaches~$\Delta$: the path turns 
around the line and goes down further along~$\Delta$ by two dodecahedra. This is 
indicated in the figure by the green cell which belongs to the path and which has 
this colour as it stands in front of~$\Pi_4$. 
\vskip 10pt
\vtop{
\vspace{-60pt}
\setbox110=\hbox{\epsfig{file=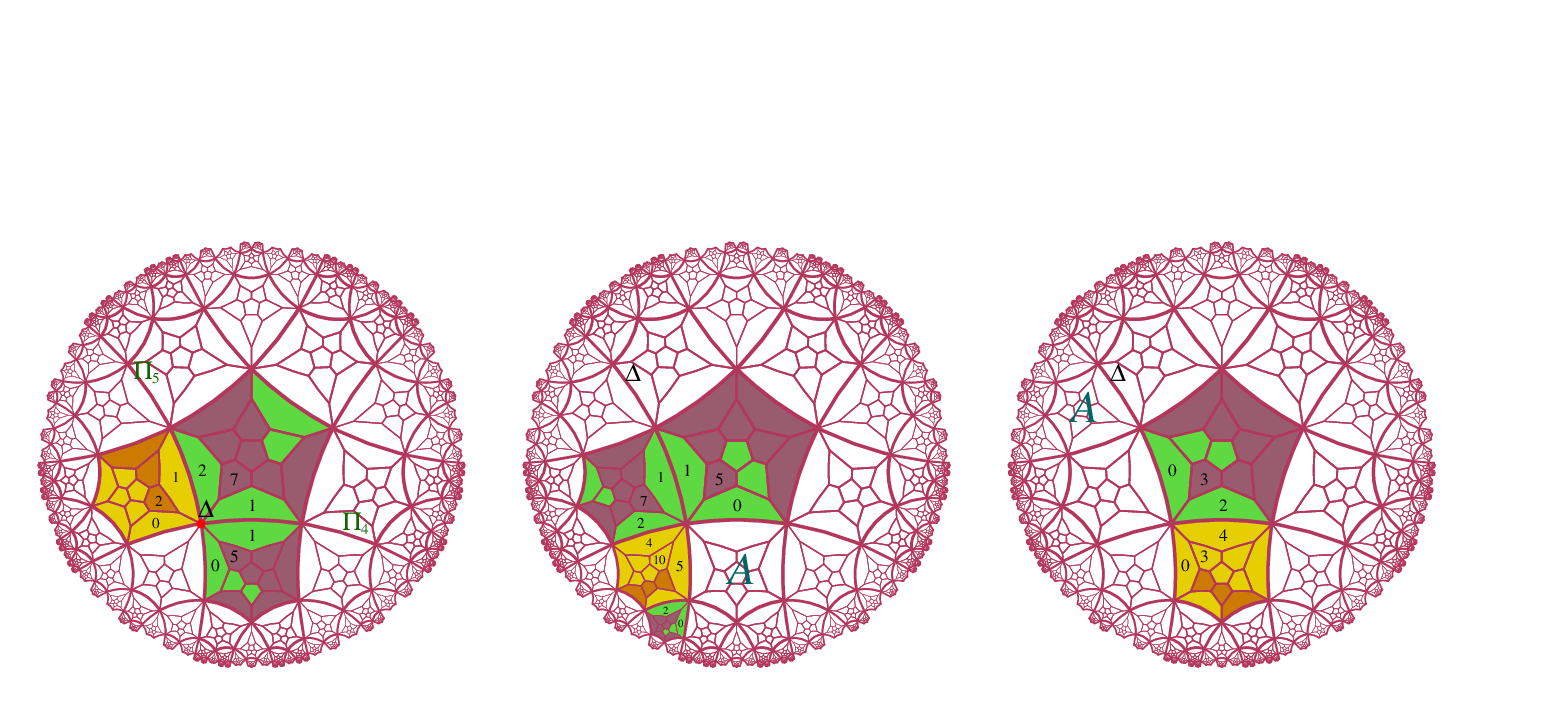,width=380pt}}
\ligne{\hfill
\PlacerEn {-175pt} {0pt} \box110
\hfill
}
\vspace{-5pt}
\begin{fig}
\label{colimacon}
\leurre
Zoom on the part of the path where the path performs a hal-turn around~$\Delta$ in
order to arrive on the appropriate farce of the cell~$13$ of the active memory switch.
On the lfet-hand side: projection onto~$\Pi_0^p$, from above. In the midle: projection
onto~$\Pi_5$: the path goes down by two dodecahedra. On the right-hand side: again
projection onto~$\Pi_5$, but from the other side as the path has passed across the plane
on its other side.
\end{fig}
}
\vskip 10pt
   The path goes down along~$\Delta$, but it does not go down until reaching cell~4.
At a distance of three tiles from cell~4, the path follows a horizontal segment 
in~$\Pi_4$ until it reaches~$\zeta$, the line of~$\Pi_4$ which passes through~$P_0^a$,
the point of~$\Pi_0^a$ which is shared by the faces~3 and~4 of cell~4{} in the 
active memory switch. Then, the path goes down along~$\zeta$, until it reaches 
the cell which is on the face~1 of cell~13. Denote this cell by~[1]$_{13}$. More 
generally, we recursively denote by $[i]_j$ the dodecahedron which is put on the 
face~$i$ of the dodecahedron denoted by~$j$ which we denote by $i_j$ if needed. 
The last index in~$j$ is a number already fixed.

In Figures~\ref{end_of_path_controlmemo} and~\ref{cell_1_13}, we have a numbering 
of the faces of the cell. Face~1 is the face which is the closest to 
cell~$[0]_{13}$, the neighbour of cell~13 which shares its face~0. In the computer 
program, cell~$[0]_{13}$ is cell~14. In Figure~\ref{cell_1_13}, it seems that the 
path goes from one face to the next one for faces~9, 8, 7 and~1. In fact, the 
path consists of the neighbours of cell~[1]$_{13}$ which are put on the just 
indicated faces together with corners which allow to go from one cell to the other. 
Indeed, the faces $9_{[1]_{13}}$ and  $8_{[1]_{13}}$ are perpendicular. 

\vskip 10pt
\vtop{
\vspace{-30pt}
\setbox110=\hbox{\epsfig{file=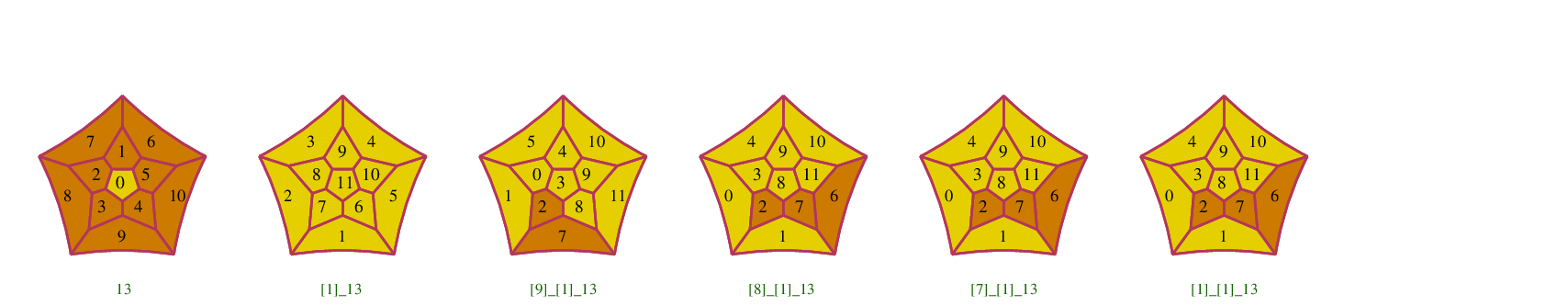,width=410pt}}
\ligne{\hfill
\PlacerEn {-175pt} {0pt} \box110
\hfill
}
\vspace{-5pt}
\begin{fig}
\label{end_of_path_controlmemo}
\leurre
The dodecahedra use of the path from~$\zeta$ to the cell~$13$ of the active memory
switch. The last five cells in the figure are straight elements: in between them, 
from the first one to the fifth one, there is a corner which may be of one or 
another of the two possible forms. Note that the last three cells have exactly the 
same pattern, rotated in the same way. Cell~$[1]_{[1]_{13}}$ is a neighbour of 
cell~$[0]_{13}$. Remember that cell~$[0]_{13}$ is cell~$14$. For all these
cells, the bottom is in contact with a black cell as this is the state of
cell~$[1]_{13}$.
\end{fig}
}

\vskip 10pt
\vtop{
\vspace{-20pt}
\setbox110=\hbox{\epsfig{file=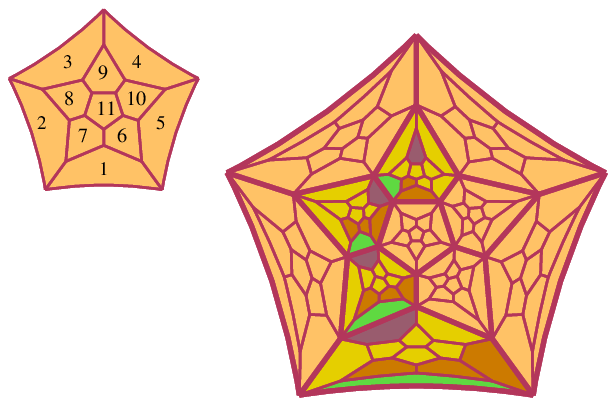,width=360pt}}
\ligne{\hfill
\PlacerEn {-175pt} {0pt} \box110
\hfill
}
\vspace{-55pt}
\begin{fig}
\label{cell_1_13}
\leurre
Illustration of the end of the path on cell~$[1]_{13}$. The green and purple faces
sharing an edge represent the corner joining the dodecahedra of the path which possess
these faces. They can also be interpreted as the entry and the exit of a cell,
in purple and green respectively.
\end{fig}
}
\vskip 10pt
Accordingly, 
their common edge, which is also shared by $[1]_{13}$ is also shared by a fourth 
dodecahedron which is a neighbour of both $[9]_{[1]_{13}}$ and $[8]_{[1]_{13}}$. 
This fourth dodecahedron can indifferently be numbered by $[a]_{[9]_{[1]_{13}}}$
or $[b]_{[8]_{[1]_{13}}}$ where~$a$ and~$b$ are appropriate faces of $[9]_{[1]_{13}}$ 
and $[8]_{[1]_{13}}$ respectively. This fourth dodecahedron has its faces~1 and~2
shared by $[9]_{[1]_{13}}$ and $[8]_{[1]_{13}}$ respectively. Which face is associated
with which of these dodecahedra is determined by the orientation of the numbering:
face~2 follows face~1 when clockwise turning around the projection of the dodecahedron
on its bottom face.

   And so, we can see that this end of the path can be represented as a word of the 
form~$(SQ)^4Q$, the penulitmate element being a corner: the one which joins the 
cell~$[1]_{[1]_{13}}$ to cell~14. Now, cell~14 is also a corner as the exit
from cell~$[1]_{[1]_{13}}$ is a face which is perpendicular to one of cell~14.
With this end of the path, we can notice that sequences of cells
of the form~$(SQ)^+$ provide us with a very flexible tool allowing to construct
a path joining any pair of cells in the dodecagrid.

   Before looking at the further connections between the active memory switch and 
the passive one, it is important to note the patterns of the straight elements 
used in the path connecting the main sensor of the passive memory switch with the 
controller of the active one. Indeed, in Figures~\ref{end_of_path_controlmemo} 
and~\ref{cell_1_13}, we can see that the just mentioned cells have two or three 
milestones. In fact, for all of them, the bottom of the cell is face~5 or~6. Now, 
these faces must be in contact with a milestone. Now, as cell~[1]$_{13}$ is always 
black, it behaves like a milestone and so, for the cells which have three milestones 
in the figures, they have in fact four of them at the requested places with respect 
to the entry and the exit. We remain with the single case of two milestones in the 
figures, it is the cell~[9]$_{[1]_{13}}$. Now, the face of this cell in contact 
with cell~$[1]_{13}$ is its face~6, so that we have three milestones on the 
following faces: 2, 6 and~7. We shall see in Section~\ref{the_rules} that with this 
pattern we can still establish rules ensuring the motion of the particle.
\vskip 10pt
   To conclude this point, we have to indicate how the tracks attached to the
active and the passive memory switches are connected. Denote by $u_a$, $a_a$
and $b_a$, $u_p$, $a_p$ and~$b_p$ the tracks meeting at the switchs with
easy notations: $u$, $a$ and~$b$ where already introduced, and the indices $a$ and~$p$
refers to the active and passive memory switches respectively. We remain with
explaining how $u_a$ and $u_p$ meet, and with the same question with $a_a$, $b_a$
and $a_p$, $b_p$ similarly. 

   In Figure~\ref{dodec_flip_flop}, we may notice that the dodecahedra belonging 
to~$u$ and~$a$ have a face belonging to the plane which supports the face~5 of the 
central cell. This means that, in the setting which we considered in this section, 
the dodecahedra belonging to~$u_a$ and~$a_a$ have a face lying on~$\Pi_5$. The same 
property occurs in the passive memory switch for the dodecahedra belonging to~$u_p$ 
and~$a_p$. At this point, we have to notice that the active and passive switches are 
put in some sense face to face: the sensors of both switches belongs to the 
intersection of two half-spaces delimited by~$\Pi_0^a$ and~$\Pi_0^p$. This means 
that, if $a_a$ denotes the right-hand side track leaving the switch in an active 
crossing, then $a_p$ denotes the left-hand side track. A similar convention is 
assumed for~$b_a$ and~$b_p$. Now, the dodecahedra belonging to~$b_a$ and~$b_p$ do 
not possess the same property with respect to~$\Pi_4$. Nevertheless, we can continue 
the tracks in such a way that, starting from a certain point, the dodecahedra of 
the tracks have a face on~$\Pi_4$. It is enough to make the track follow
a horizontal segment starting from the third cell after the central one, until the
track meets the plane.
 
   Now, assuming this point, we can define the route followed by the tracks designed
by the same first letter as a route similar to the one used for the bridges. In the
planes~$\Pi_4$ and~$\Pi_5$, we can define the levels of four Fibonacci trees rooted 
at~$A$ and take one of them which will guide the tracks, call this level the 
{\bf border}. The border meet the plane~$P_0$ at some point. This means that in the 
plane~$\Pi_0$, when the tracks meet the border, they are separated. One way go up to 
the plane~$\Pi_0^p$ and, there, it follows the plane~$\Pi_4$ or~$\Pi_5$, depending 
on which track we consider, until the central cell of the passive switch is met. The 
other track goes down along the border until it meets the plane~$\Pi_0^a$ and from 
there, it goes to the central cell of the active switch. The needed length for the 
radius of the border is defined by the number of cells needed by the path from the 
controller of the passive switch until that of the active switch. A radius of 
10~tiles is more than enough for our purpose, as can be seen from 
Figures~\ref{colimacon}, \ref{end_of_path_controlmemo} and~\ref{cell_1_13}.

   From this, we have to notice that outside the level defined by the border, the 
tracks $u_a$, $u_p$, $a_a$, $a_p$, $b_a$ and~$b_p$ need not to go on 
following~$\Pi_4$ or~$\Pi_5$.
\vskip 10pt
   We have now presented all the elements of the scenario defining the simulation
of our railway circuit. We can now turn to the study of the rules and the computer
program.
   
\vskip 10pt
\section{The rules and the computer program}
\label{the_rules}
   In this section, we shall follow the guidelines of the previous section and, at
each step, we shall formulate the rules corresponding to the considered part of 
the scenario. In each step, we shall have two kinds of rules: {\bf conservative} 
rules and {\bf motion} ones. Conservative rules are characterized by the fact that
the new state is the same as the current one. In motion rules, the new state is 
different from the current one. This syntactic difference expresses the presence or 
not of the particle nearby the cell or within it. However, in the motion rules,
we shall include rules which look like conservative. As these rules involve the
particle, they are considered as motion rule too. We shall have an illustration of
this point in the flip-flop switch for instance.

   Here, we shall take the same format for the rules as the one used 
in~\cite{mmarXiv3}. The rule will be presented as a word~$\eta^0\eta_0..\eta_11\eta^1$ 
of length~14, where $\eta^0$ is the current state of the cell, $\eta_i$ is the cell of
the $i^{\rm th}$~neighbour of the cell and $\eta^1$ is the new state of the cell after 
the application of the rule. In this paper, we shall display the rules according
to the display used in the file treated by the computer program. As can be seen, 
the notations are straightforward from the just indicated formalism.

   Here we present this new display by Table~\ref{basic_cons} which defines
a few samples of the basic conservative rules defined by the following principle: 
when a cell has at most three black neighbours, its state remains unchanged.
In all the tables where the rules are displayed, we also display a numbering of
the faces which allows the reader to easily make the correspondence with the
states of a face.
 
\subsection{The rules for the tracks}
\label{track_rules}

   For the rules concerning the tracks, we remember that the patterns are the same
than those of~\cite{mmarXiv3}. Now, the rules are different. We remind that for the
straight element, the faces which are shared with a milestone are faces~2, 5, 6 
and~7 with, in this context, the entry through cell~3, 4, 8 or~10 and the exit 
through face~1. Most often, the bottom of the cell is face~0 or face~5. But, 
in principle, other rotations are possible. For the straight element, the rules 
are given
by Table~\ref{rules_straight}. 

\ligne{\hfill}
\vtop{
\vspace{-15pt}
\begin{tab}\label{basic_cons}
\leurre
The basic conservative rules: the state of a cell which has at most three black neighbours
remains unchanged.
\end{tab}
\vspace{-12pt}
\grostrait
\vskip3pt
{\obeylines
\obeyspaces\global\let =\ \ttix\parskip=-2pt
\vskip3pt
--adress  -1   0   1   2   3   4   5   6   7   8   9  10  11  12
\vskip3pt
(0)        W   W   W   W   W   W   W   W   W   W   W   W   W   W
(0)        B   W   W   W   W   W   W   W   W   W   W   W   W   B
(0)        B   B   W   W   W   W   W   W   W   W   W   W   W   B
(0)        W   B   W   W   W   W   W   W   W   W   W   W   W   W
(0)        W   B   B   W   W   W   W   W   W   W   W   W   W   W
(0)        B   B   B   W   W   W   W   W   W   W   W   W   W   B
(0)        B   B   B   B   W   W   W   W   W   W   W   W   W   B
}
\demitrait
}

\vskip 10pt
\vtop{
\begin{tab}\label{rules_straight}
\leurre
The rules for the straight elements. The conservative rules are labelled 
with~$0$. Note that the motion is always in the same direction: the exit
is through face~$1$. Also note the special element with three milestones
only.
\end{tab}
\vspace{-12pt}
\grostrait
\vskip3pt
{\obeylines
\obeyspaces\global\let =\ \ttix\parskip=-2pt
\vskip3pt
--adress  -1   0   1   2   3   4   5   6   7   8   9  10  11  12
\vskip3pt
-- straight element, direct track:
--
(0)        W   W   W   B   W   W   B   B   B   W   W   W   W   W
--
--   presence of the particle:
--
(1-3)      W   W   W   B   B   W   B   B   B   W   W   W   W   B
(1-4)      W   W   W   B   W   B   B   B   B   W   W   W   W   B
(1-10)     W   W   W   B   W   W   B   B   B   W   W   B   W   B
(1-8)      W   W   W   B   W   W   B   B   B   B   W   W   W   B
(2)        B   W   W   B   W   W   B   B   B   W   W   W   W   W
(3)        W   W   B   B   W   W   B   B   B   W   W   W   W   W
--
\ifnum1=0 {
-- straight element, return track:
--
--adresse -1   0   1   2   3   4   5   6   7   8   9  10  11  12
--
(0)        W   B   W   B   W   W   B   W   B   W   W   W   W   W
--
--   presence of the particle:
--
(1-4)      W   B   W   B   W   B   B   W   B   W   W   W   W   B
(1-3)      W   B   W   B   B   W   B   W   B   W   W   W   W   B
(2)        B   B   W   B   W   W   B   W   B   W   W   W   W   W
(3)        W   B   B   B   W   W   B   W   B   W   W   W   W   W
--
\} \fi
-- special straight element, memory switch:
--
(0)        W   W   W   B   W   W   W   B   B   W   W   W   W   W
(1)        W   W   W   B   W   B   W   B   B   W   W   W   W   B
(2)        B   W   W   B   W   W   W   B   B   W   W   W   W   W
(3)        W   W   B   B   W   W   W   B   B   W   W   W   W   W
}
\demitrait
}
\vskip 10pt
   In Table~\ref{rules_straight}, we can notice that the rules labelled with~$(0)$
are conservative and that the others are concerned with the motion of the particle.

   An interesting feature of the rules given in Table~\ref{rules_straight} is
that the rules associated with the return track are indeed rotated forms of the
other rules. This comes from the symmetry of the milestone pattern with respect to
a certain plane reflection leaving the dodecahedron invariant: it is namely
the bisector of the angles defined by the two 'inner' faces shared with a milestone. 

\vskip 10pt
\vtop{
\vspace{-15pt}
\begin{tab}\label{rules_corner}
\leurre
The rules for the four kinds of corners.
\end{tab}
\vspace{-12pt}
\grostrait
\vskip3pt
{\obeylines
\obeyspaces\global\let =\ \ttix\parskip=-2pt
\vskip3pt
--adress  -1   0   1   2   3   4   5   6   7   8   9  10  11  12
\vskip3pt
--
-- corners:
--
(0)        W   W   W   W   B   W   B   B   B   B   W   B   B   W
(0)        W   W   W   W   B   W   B   B   B   B   B   W   B   W
--
--   version with B upon face 0:
--
(0)        W   B   W   W   B   W   B   B   B   B   B   W   B   W
(0)        W   B   W   W   B   W   B   B   B   B   W   B   B   W
--
--   when the particle is nearby:
--
--     direction :  1 -> 2
--
(1)        W   W   B   W   B   W   B   B   B   B   W   B   B   B
(2)        B   W   W   W   B   W   B   B   B   B   W   B   B   W
(3)        W   W   W   B   B   W   B   B   B   B   W   B   B   W
--
--         version with B upon face 0:
--
(1)        W   B   B   W   B   W   B   B   B   B   W   B   B   B
(2)        B   B   W   W   B   W   B   B   B   B   W   B   B   W
(3)        W   B   W   B   B   W   B   B   B   B   W   B   B   W
--
--     direction :  2 -> 1
--
(1)        W   W   W   B   B   W   B   B   B   B   B   W   B   B
(2)        B   W   W   W   B   W   B   B   B   B   B   W   B   W
--
(3)        W   W   B   W   B   W   B   B   B   B   B   W   B   W
--
--         version with B upon face 0:
--
(1)        W   B   W   B   B   W   B   B   B   B   B   W   B   B
(2)        B   B   W   W   B   W   B   B   B   B   B   W   B   W
(3)        W   B   B   W   B   W   B   B   B   B   B   W   B   W
--
--adress  -1   0   1   2   3   4   5   6   7   8   9  10  11  12
}
\demitrait
}
\vskip 10pt
At last, we have also the rules for the special motion of the temporary particle
conveying the signal to the controller of the active memory switch. Here there are
only three milestones put onto face~2, 6 and~7, the exit still being face~1 and the
entry being face~4 or~3.

   As indicated in Subsection~\ref{implement_tracks}, there is another important
element for constructing the tracks: the corners. In Subsection~\ref{implement_tracks},
the corners are given in two versions which are symmetric images of one another in a 
reflection through a plane. And so, these two versions are not rotated images of each 
other. In one version, if the entry is face~1 and the exit is face~2, then the 
milestones are put onto faces~2, 3, 5, 6, 7, 8, 10 and~11. If we consider the 
bisector of the angle defined by the planes of faces~1 and~2, the corner is almost 
symmetric with respect to the reflection in this plane, but faces~9 and~10 are 
exchanged by this reflection and one face is white while the other is black. We 
take advantage of this situation to consider the symmetric image as defining 
the reverse motion: the entry is now face~2 and the exit is face~1.

   But later, especially for the passive memory switch and for defining the path
from its sensor to the controller of the active memory switch, we have two introduce
another version of both corners: the new pattern consists in putting an additional 
milestone on face~0. The necessity of this can be remarked from
Figure~\ref{colimacon}: it is clear that in several cases, face~0 of a corner
must be shared by a black cell. Table~\ref{rules_corner} shows the rules for all 
these versions of the corners.

\subsection{The rules for the flip-flop switch}
\label{flip_flop_rules}

   Before turning to the flip-flop switch, it sould be remarked that, from the 
structure of a straight element, there is no need to introduce additional rules 
for the working of the fixed switch: the rules defined in 
Subsection~\ref{track_rules} are enough to ensure the correct working of this
type of switches.

   Now, let us have a look to the flip-flop switch, which is a purely
active switch. As mentioned in Subsection~\ref{flip-flop}, it is needed to
introduce new patterns.

   The first one deals with the central cell. As noticed in 
Subsection~\ref{flip-flop}, the central cell has a special pattern. Consider
a numbering where the track~$u$ abuts face~1 while $a$~and~$b$ abut faces~4 and~3
repsectively, the face in contact with~$\Pi_0$ being face~0. Then the milestones 
are put onto faces 2, 5, 8, 10 and~11, see Figure~\ref{dodec_flip_flop}.

   The other patterns are a consequence of the mechanism introduced for the
working of the switch as a flip-flop one. In Figure~\ref{control_flip_flop},
we have seen the three patterns used for this purpose with the appropriate
numbering of each new cell. In these numberings, cells~11 and~12 have only three
milestones which are on faces~1, 6 and~7. Note that the conservative rules
associated to these cells are not rotated forms of the rules given in
Table~\ref{basic_cons}. Cell~13 has a more complex pattern: its milestones
are on faces~1, 6, 7, 8~and~10.

   However, we can see on the conservative rules of cell~13 that exactly 6~black 
cells occur in both rules. This comes from the fact cell~13 is in contact with
both cells~[9]$_5$ and~[9]$_8$. Always one exactly of these two cells is black, 
but which one depends on the motion of the particle. In fact, in the configuration
of a flip-flop switch, outside the cells of the track, exactly three cells among
all heir neighbours can change their state: cells~11, 12 and~13. We have seen 
the 
\ligne{\hfill}
\vtop{
\vspace{-15pt}
\begin{tab}\label{rules_flip_flop}
\leurre
The rules for the flip-flop.
\end{tab}
\vspace{-12pt}
\grostrait
\vskip3pt
{\obeylines
\obeyspaces\global\let =\ \ttix\parskip=-2pt
\vskip3pt
--adress  -1   0   1   2   3   4   5   6   7   8   9  10  11  12
\vskip3pt
--  central cell:
(0)        W   W   W   B   W   W   B   W   W   B   W   B   B   W
(0)        W   W   W   B   W   W   B   W   W   B   B   B   B   W
--
--    when it is crossed by the particle:
--
(1)        W   W   B   B   W   W   B   W   W   B   W   B   B   B
(2)        B   W   W   B   W   W   B   W   W   B   W   B   B   W
(3-3)      W   W   W   B   B   W   B   W   W   B   W   B   B   W
(3-4)      W   W   W   B   W   B   B   W   W   B   W   B   B   W
--  
--  cells 5 and 8:
--
(0)        W   W   W   B   W   W   B   B   B   W   B   W   W   W
--
--    when a particle passes nearby:
--
(1)        W   W   W   B   B   W   B   B   B   W   B   W   W   W
(1)        W   W   W   B   W   B   B   B   B   W   B   W   W   W
(0)        W   W   B   B   W   W   B   B   B   W   B   W   W   W
--
-- cell 13, the controller:
--
(0)        W   W   B   W   W   B   W   B   B   B   W   B   W   W
(0)        W   W   B   W   B   W   W   B   B   B   W   B   W   W
--
--   activation of cell 13 (the particle is seen through face 0):
--
(1-3)      W   B   B   W   W   B   W   B   B   B   W   B   W   B
(1-4)      W   B   B   W   B   W   W   B   B   B   W   B   W   B
(2)        B   W   B   W   W   B   W   B   B   B   W   B   W   W
(2)        B   W   B   W   B   W   W   B   B   B   W   B   W   W
--
-- cells 11 and 12
--
(0)        W   W   B   W   W   W   W   B   B   W   W   W   W   W
(0)        B   W   B   W   W   W   W   B   B   W   W   W   W   B
--
--    change of state in cells 11 and 12:
--
(1-W)      W   B   B   W   W   B   W   B   B   W   W   W   W   B
(1-W)      W   B   B   W   B   W   W   B   B   W   W   W   W   B
(1-B)      B   W   B   W   W   B   W   B   B   W   W   W   W   W
(1-B)      B   W   B   W   B   W   W   B   B   W   W   W   W   W
--
--adresse -1   0   1   2   3   4   5   6   7   8   9  10  11  12
}
\demitrait
}
\vskip 10pt
\noindent
reason of these changes and what is their meaning in Section~\ref{scenario}.
We can see in Table~\ref{rules_flip_flop} how the rules implement the scenario,
very directly. In particular, we can see that the presence of a black neighbour
on the face~9 of cell~5 or cell~8 is enough to prevent the particle to enter
the track which follows cell~5 or cell~8 respectively. When cell~11 or~12
is white, cell~5 or~8 respectively behaves like an ordinary straight element
and so, the particle can cross the cell and go on along the corresponding track.
We can also see that the flash of cell~13 is triggered by the occurrence
of two black neighbours: one through face~3 or~4 and the other through face~0.
The first one is the marker of the non selected track and the second is the
particle itself which is passing through cell~4.

\subsection{The rules for the memory switch}

   Now, we come to the rules needed by the memory switch. We know that this
switch is split into two components: the passive memory switch which deals with 
passive crossings only and the active memory switch which deals with active
passages only.

   The rules concerning the passive memory switch are
displayed by Table~\ref{rules_passive_memo}. We can see that there is no special 
rule for the central cell: it is a straight element with face~0 on~$\Pi_0$,
exactly as in a fixed switch.

   Now, we can seen that, as explained in Subsection~\ref{memory}, the cells~11, 12
and~13 of the passive memory switch are very different from those of a flip-flop.

   In particular, for the passive memory switch, cells~11 and~12 have three milestones
two, but they are placed on faces~1, 8 and~10, which makes this pattern different
from that of the cells numbered in the same way for the flip-flop switch, even under
any positive displacement keeping a dodecahedron globally invariant. This is entailed
by the fact that cells~11 and~12 do not react in the same way as their homonyms.
In the flip-flop switch, cells~11 and~12 see a neighbour of the central cell: cell~13
which is cell~[9]$_4$. Here, this is not the case. Cells~11 and~12 can see cell~13
only on which they are put. They can see a neighbour of cell~5 or~8 respectively, but
this neighbour is always white. 

   Now, cell~13 is also very different: it has 5~milestones on faces~0, 1, 6, 7 
and~11. Moreover, one of its two neighbours seen through faces~8 and~10, is black
and the other is white. These neighbours are namely cell~11 and~12, on faces~10
and~8 respectively. The conservative rules reflect this structure. By its
position on cell~13, The black cell indicates the side which is the non selected 
track, this is why we say that these cells are {\bf markers}. The motion rules 
show that cell~13 becomes black as soon as the particle is seen on the track 
which is on the same side as its black marker. Now, cell~13 sees the particle before
it enters the central cell while, in the case of the flip-flop, it can see it one
time later only: this is due to to the position of the central cell. 
\vskip 10pt
   Table~\ref{rules_active_memo} indicate the rules for the active memory switch.
This time, what is common with the flip-flop memory switch also applies here. This 
is the case, in particular, for cells~11 and~12 which have the same pattern as
in the case of the flip-flop switch. Note too that here, they have the same 
position as in this latter switch. They play exactly the same role as in the flip-flop
switch. However, their change of state does not occur in the same way as in the
flip-flop switch, where the change is caused by the presence of the particle
in the central cell. In the active switch, the change occurs much later than 
the time when the particle visits the central cell. However, the particle is not
present neither in cell~5 nor in cell~8 when cell~13 flashes for cells~11 and~12
to change their state. This makes it necessary to append three new rules only:
the corresponding rules when one of the cells is black are already present in
Table~\ref{rules_flip_flop} and, accordingly, they also apply here.     
\vskip 10pt
\vtop{
\vspace{-15pt}
\begin{tab}\label{rules_passive_memo}
\leurre
The rules for the passive memory switch.
\end{tab}
\vspace{-12pt}
\grostrait
\vskip3pt
{\obeylines
\obeyspaces\global\let =\ \ttix\parskip=-2pt
\vskip3pt
--adress  -1   0   1   2   3   4   5   6   7   8   9  10  11  12
\vskip3pt
--
--   cell 13:
--
(0)        W   B   B   W   W   W   W   B   B   W   W   B   B   W
(0)        W   B   B   W   W   W   W   B   B   B   W   W   B   W
--
--      when the particle passes, seen through face 3 or 4,
--         it flashes, if cell 8 or 10 respectively is black:
--
(1)        W   B   B   W   W   B   W   B   B   W   W   B   B   B
(1)        W   B   B   W   B   W   W   B   B   B   W   W   B   B
(2)        B   B   B   W   W   W   W   B   B   W   W   B   B   W
(2)        B   B   B   W   W   W   W   B   B   B   W   W   B   W
--
--         it does not flash, if cell 8 or 10 respectively is white:
--
(0)        W   B   B   W   W   B   W   B   B   B   W   W   B   W
(0)        W   B   B   W   B   W   W   B   B   W   W   B   B   W
--
--      when cell [9]\_13 flashes after the flash of cell 13: 
--
(0)        W   B   B   W   W   W   W   B   B   B   B   W   B   W
(0)        W   B   B   W   W   W   W   B   B   W   B   B   B   W
--
--   cells 11 and 12:
--
(0)        W   W   B   W   W   W   W   W   W   B   W   B   W   W
(0)        B   W   B   W   W   W   W   W   W   B   W   B   W   B
(0)        W   B   B   W   W   W   W   W   W   B   W   B   W   B
(0)        B   B   B   W   W   W   W   W   W   B   W   B   W   W
--
--      effect on cells 5 and 8 when cell 13 flashes
--
(0)        W   B   W   B   W   W   B   B   B   W   W   W   W   W
(0)        W   B   B   B   W   W   B   B   B   W   W   W   W   W
--
--adress  -1   0   1   2   3   4   5   6   7   8   9  10  11  12
}
\demitrait
}
\vskip 10pt
   Now, cell~13 has also a different working than its homonym of the flip-flop
switch. The main difference is that here, cell~13 remains white just after 
the visit of the particle. Accordingly, it is triggered to flash in a different
way. We can see the rules corresponding to this new situation in 
Table~\ref{rules_active_memo}.
\vskip 10pt
\vtop{
\vspace{-15pt}
\begin{tab}\label{rules_active_memo}
\leurre
The rules for the active memory switch.
\end{tab}
\vspace{-12pt}
\grostrait
\vskip3pt
{\obeylines
\obeyspaces\global\let =\ \ttix\parskip=-2pt
\vskip3pt
--adress  -1   0   1   2   3   4   5   6   7   8   9  10  11  12
\vskip3pt
--
--   cell 13:
--
(0)        W   W   B   B   B   B   B   B   W   B   B   B   W   W
(0)        W   W   B   B   B   B   B   W   B   B   B   B   W   W
--
--      no change when the particle passes under face 11:
--
(0)        W   W   B   B   B   B   B   B   W   B   B   B   B   W
(0)        W   W   B   B   B   B   B   W   B   B   B   B   B   W
--
(1)        W   B   B   B   B   B   B   W   B   B   B   B   W   B
(1)        W   B   B   B   B   B   B   B   W   B   B   B   W   B
(2)        B   W   B   B   B   B   B   W   B   B   B   B   W   W
(2)        B   W   B   B   B   B   B   B   W   B   B   B   W   W
--
--    cells 11 and 12{} in the active memory switch:
--
(2)        W   B   B   W   W   W   W   B   B   W   W   W   W   W
(3)        W   W   B   W   W   B   W   B   B   W   W   W   W   B
(3)        W   W   B   W   B   W   W   B   B   W   W   W   W   B
}
\demitrait
}
\vskip 10pt

\subsection{A word about the computer program}

   The computer program is based on the same one which was used to check the
rules of the weakly universal cellular automaton on the dodecagrid with three
states which was used in~\cite{mmarXiv3}. However, it was adapted to this automaton
in the part computing the initial configuration and, also in the main function
as the control steps are a bit different from those of the previous program.
As already mentioned, all the traces given in Section~\ref{scenario} are
taken from a general trace computed by the program.

   Now, from all the features proved in Section~\ref{scenario} and from
the computations performed in this one, we can conclude that the proof of
Theorem~\ref{weakuniv2} is now complete.

\section{Conclusion}
\label{conclusion}

   With this result, we reached the frontier between decidability and weak universality
for cellular automata in hyperbolic spaces: starting from 2~states there are weakly 
universal such cellular automata, with 1~state, there are none, which is trivial.
Moreover, the set of cells run over by the particle is a true spatial stucture.
We can see that the third dimension is much more used in this implementation than
in the one considered in~\cite{mmarXiv3}.
 
   What can be done further?

   In fact, the question is not yet completely closed. In~\cite{mmarXiv1DinkD}
we proved that it is possible to implement a $1D$-cellular automaton with $n$~states
into the pentagrid, the heptagrid and the dodecagrid and also a whole family of
tilings of the hyperbolci planei with the same number of states. For the
pentagrid, it was needed to append an additional condition which is satisfied,
in particular, by the elementary cellular automaton defined by rule~110. 
Consequently, as stated in~\cite{mmarXiv1DinkD}:

\begin{thm}\label{weakuniv2math} \rm{(M. Margenstern, \cite{mmarXiv1DinkD})}.
There is a weakly universal rotation invariant cellular automaton in the pentagrid,
in the heptagrid and in the dodecagrid.
\end{thm} 

   However, this is a general theorem based on the very complicate proof of a deep
result involving a number of computations in comparison with which those of
this paper are quasi-nothing. Also, the implementation provides a structure
which is basically a linear one. This is why, it seems to me that the construction of
the present paper is worth of interest: it is a truly spatial construction. Moreover,
the construction is very elementary.

   Now, there are still a few questions. What can be done in the plane with a true
planar construction? At the present moment, the smaller number of states is~4,
in the heptagrid, see~\cite{mmTCS4}, while it is 9~in the pentagrid,
see~\cite{mmsyPPL}. And so, there is some definite effort 
before closing this question.

\end{document}